\documentclass[review, 12pt, a4paper]{elsarticle}
\usepackage[margin=3cm]{geometry}
\usepackage{amsfonts,amsmath,amsthm,amssymb}
\usepackage{tabularx}
\usepackage{multirow}
\usepackage[export]{adjustbox}
\usepackage[labelfont=bf,justification=raggedright,singlelinecheck=false]{caption}
\usepackage{caption}
\usepackage{xcolor}
\usepackage{bm,bbm, dsfont}
\usepackage{graphicx,xspace,epsfig,syntonly,dsfont}
\usepackage{subcaption}
\usepackage{lipsum}
\usepackage{siunitx}
\usepackage[stable]{footmisc}
\usepackage{url}
\usepackage{varwidth}
\usepackage{xfrac}
\usepackage{optidef}
\usepackage{cleveref}
\usepackage{booktabs}
\usepackage{makecell}
\usepackage{array}
\newcolumntype{?}{!{\vrule width 1pt}}

\usepackage{placeins}

\usepackage{nomencl}
\makenomenclature

\usepackage{etoolbox}
\renewcommand\nomgroup[1]{%
  \item[\bfseries
  \ifstrequal{#1}{I}{Sets and Indices}{%
  \ifstrequal{#1}{P}{Parameters}{%
  \ifstrequal{#1}{V}{Decision Variables}{%
  \ifstrequal{#1}{D}{Uncertain Parameters}{%
  \ifstrequal{#1}{O}{Objective Values}{%
  }}}}}%
]}

\biboptions{sort&compress}
\allowdisplaybreaks
\begin{document}
\begin{frontmatter}

\title{Resilient Grid Hardening against Multiple Hazards:\\ An Adaptive Two-Stage Stochastic Optimization Approach}

\author{Sifat Chowdhury, Yihsu Chen and Yu Zhang \corref{cor1}}
\cortext[cor1]{Corresponding author, email: zhangy@ucsc.edu}

\address{Department of Electrical and Computer Engineering\\ University of California, Santa Cruz}


\begin{abstract}
The growing prevalence of extreme weather events driven by climate change poses significant challenges to power system resilience. Infrastructure damage and prolonged power outages highlight the urgent need for effective grid-hardening strategies. While some measures provide long-term protection against specific hazards, they can become counterproductive under conflicting threats. In this work, we develop an adaptive two-stage stochastic optimization framework to support dynamic decision-making for hardening critical grid components under multiple hazard exposures. Unlike traditional approaches, our model adapts to evolving climate conditions, enabling more resilient investment strategies. Furthermore, we integrate long-term (undergrounding) and short-term (vegetation management) hardening actions to jointly minimize total system costs. Extensive simulation results validate the effectiveness of the proposed framework in reducing outage and repair costs while enhancing the adaptability and robustness of grid infrastructure planning.
\end{abstract}

\end{frontmatter}

\newpage

\mbox{}
\nomenclature[I]{\(i,j\)}{Bus indices}
\nomenclature[I]{\(\mathcal{N}^{\text{L}/ \text{MT}}\)}{Set of load buses and microturbine (MT) generator buses, respectively}
\nomenclature[I]{\(\tilde{\mathcal{T}};t \)}{Set of years in the investment period $\tilde{\mathcal{T}}\triangleq \{1,2,\dots,T\}$ and yearly index}
\nomenclature[I]{\(e\)}{Index denoting extreme weather event including earthquake (EQ), high wind (W), and fallen trees (VM)}
\nomenclature[I]{\(\mathcal{L}; l\)}{Set of all distribution lines and line index}
\nomenclature[I]{\(\mathcal{H};h\)}{Set of 24 hours in a day $\mathcal{H}\triangleq \{1,2,\dots,24\}$ and hourly index}

\nomenclature[O]{\(v^{\text{TS}}\); \(v^{\text{ATS}}\)}{Optimal objective value of the two-stage and adaptive two-stage problem, respectively (million dollars)}

\nomenclature[V]{\(\alpha_{ln}\)}{Binary variable indicating undergrounding decision for line $l$ at node $n$: 1 to be undergrounded, 0 otherwise}
\nomenclature[V]{\(\beta_{ln}\)}{Variable indicating the percentage of vegetation around line $l$ at node $n$ that needs to be managed}
\nomenclature[V]{\(r_{lt}\)}{Binary variable indicating if the decision $\{\alpha_{ln}\}$ are revised at node $n$ $\in S_{t}$: 1 if revised, 0 otherwise}
\nomenclature[V]{\(P_{i}; Q_{i}\)}{Active and reactive power injection at bus i (\si{MW; MVar})}
\nomenclature[V]{\(P_{l}; Q_{l}\)}{Active and reactive power flow over branch $l$ (\si{MW; MVar})}
\nomenclature[V]{\(\ell_{l}; v_{i}\)}{Squared magnitude of current flowing over line $l$, and squared magnitude of voltage at bus $i$}

\nomenclature[P]{\(\text{IC}^{\text{UG}}_{ln}; \text{IC}^{\text{VM}}_{ln}\)}{Cost of undergrounding and VM for line $l$ at node $n$ (\si{\$/mile})}
\nomenclature[P]{\(\delta_{ln}^{\text{VM}}\)}{Criticality factor of line $l$ at node $n$ requiring VM}
\nomenclature[P]{\(\delta_{ln}^{\text{UG}}\)}{Ratio of the future projected outage and repair cost of earthquake and wind of line $l$ at node $n$}
\nomenclature[P]{\(B^{\text{UG}}; B^{\text{VM}}\)}{Total budget for undergrounding (in million dollars) and VM (\si{\$})}
\nomenclature[P]{\(p_n\)}{Probability of node $n$}
\nomenclature[P]{\(P^{\text{L,e}}_{i,h}\)}{Active power load demand during event $e$, at bus $i$ and time $h$ (\si{MW})}
\nomenclature[P]{\(u_{l}\)}{Binary indicator variable representing line status: 1 if in service, 0 otherwise}
\nomenclature[P]{\(\text{C}^{\text{ENS}}\)}{Cost of energy not served (\si{\$/ MWh})}
\nomenclature[P]{\(\text{C}_{l}^{\text{LS,e}}\)}{Load shedding cost for  outaged line $l$ during event $e$ (\si{\$/ day})}

\nomenclature[P]{\(\text{OC}_{ln}^{\text{W/EQ/VM}}\)}{Outage cost of line $l$ at node $n$ due to wind/earthquake/fallen trees (\si{\$})}
\nomenclature[P]{\(\text{RC}_{ln}^{\text{W/EQ/VM}}\)}{Repair cost of line $l$ at node $n$ due to wind/earthquake/fallen trees (\si{\$})}
\nomenclature[P]{\(\rho_{ln}\)}{Observed 
 percentage of vegetation in line $l$ at node $n$ }
 \nomenclature[P]{\(\mu_{l}\)}{Length of line $l$ (\si{mile})}
 \nomenclature[P]{\(f; d\)}{Inflation rate; discount rate}
\nomenclature[P]{\(N^\text{UG}_{n}\)}{Maximum number of lines that can be undergrounded in node $n$ }
\nomenclature[P]{\(\varrho_{ln}^{\text{W}/\text{EQ}/\text{VM}}\)}{Failure probability of line $l$ at node $n$ due to wind/earthquake/fallen trees}

\nomenclature[D]{\(\xi_{n}^{\text{W}/\text{EQ}/\text{VM}}\)}{Outage duration  (days) due to high wind, earthquake, and no VM at node $n$, respectively}
\nomenclature[D]{\(N_{n}^{\text{W}/\text{EQ}/\text{VM}}\)}{Projected number of high wind, earthquake, and no VM events in the year corresponding to node $n$, respectively}
\printnomenclature[0.85in]
\vspace{2em}

\section{Introduction}
Rising global temperatures have intensified extreme weather events, including storms, hurricanes, heat waves, earthquakes, etc. These events increasingly damage power system infrastructure, such as downed transmission lines, flooded substations, and other critical components, resulting in prolonged and widespread outages. Restoring power under such conditions requires substantial time and resources, causing disruptions to households, businesses, and essential services \cite{cc_influence_review}. 
In 2022 alone, the country experienced 18 major weather events, each causing economic losses exceeding 1 billion US dollars \cite{financial_loss}. As climate change continues to increase the frequency, intensity, and duration of these disasters, the associated financial and societal impacts are expected to escalate. These challenges highlight the urgent need for investment in grid hardening strategies that improve the resilience of critical infrastructure and ensure the reliable delivery of electricity during and after extreme weather conditions.

\subsection{Literature review}
 
Various natural disasters disrupt the normal operation of the power grid. Structural damage caused by high winds is a leading cause of power outages. 
Poudyal et al. developed a probabilistic weather-impact model to analyze and quantify the spatiotemporal effects of hurricanes on the grid infrastructure \cite{hurricane1}. In another study, the authors in \cite{hurricane2} proposed a combined operational performance index and weather-based failure rate index to assess the vulnerability of transmission lines under strong wind conditions. Earthquakes, among the most unpredictable natural disasters, can cause extensive damage to substations and transmission and distribution networks. They often result in structural collapse and disrupted power delivery due to equipment malfunctions. Analytical models have been developed to characterize seismic hazards and assess the vulnerability of power distribution system components \cite{tri-levelopt}, \cite{eq3}. In addition, researchers have examined large-scale service disruptions caused by other natural hazards such as floods \cite{flood2} and wildfires \cite{wildfire3}.


Resilience enhancement strategies aimed at protecting power systems from severe weather events encompass a variety of grid hardening approaches, including the reinforcement and upgrading of structural equipment \cite{tri-levelopt}, \cite{UG1}, the deployment of distributed energy resources \cite{ess_dg_allocation}, and vegetation management. The choice of an appropriate hardening strategy depends on the specific hazards to which a region is most vulnerable. For instance, placing overhead power lines underground is one of the most effective methods for protecting infrastructure in areas prone to high winds. This approach is extensively studied in \cite{UG1}. Vegetation management, such as trimming tree branches near power lines, also reduces outage risks, particularly those caused by falling trees or limbs contacting overhead lines. The study in \cite{vm} examines the relationship between outage frequency and timely vegetation management in distribution systems, while \cite{veg_risk} develops spatially explicit vegetation risk models with fine resolution to guide such programs. 
However, a key limitation of many of these studies is their narrow focus on either a single hardening strategy or a single type of hazard. Although some research considers combinations of measures, such as undergrounding power lines and deploying mobile generators \cite{multi2}, or evaluates multiple strategies for typhoon resilience \cite{AE_multiple}, the compounding risks posed by multiple hazards are often neglected. In practice, many regions face concurrent threats, such as both high winds and seismic activity, yet the interplay between these hazards remains insufficiently addressed in the existing literature.

Given the substantial financial losses associated with power outages and the significant capital investment required to harden long-lived grid assets, it is critical to make optimal deployment decisions informed by long-term cost-benefit analyses. Over the years, researchers have explored a range of optimization techniques to support this decision-making process, with linear programming being among the most widely used methods \cite{ess_allocation}. Probabilistic approaches have also been examined, including two-stage stochastic optimization \cite{multi-strategies}. In the context of robust decision-making, \cite{robust} investigates expansion planning for multi-energy distribution networks while \cite{DRO2} employs a distributionally robust optimization technique. 
Several variants of these frameworks have been developed, including stochastic robust optimization \cite{UG1}, stochastic distributionally robust models \cite{stochastic_DR}, tri-level defender-attacker-defender formulations \cite{tri-levelopt}.

A notable limitation of many of these studies is the assumption of static decision variables over extended planning horizons. Relying on a fixed set of early-stage hardening decisions, as is typical in traditional two-stage optimization, without accounting for evolving system conditions and uncertainties, may result in suboptimal and costly outcomes. This challenge can be addressed by incorporating multiple decision stages, as demonstrated in \cite{multistage}, where the decision maker updates decisions at each stage based on the current state of the system. However, the irreversible nature of certain hardening actions, such as undergrounding power lines, limits the feasibility of frequent adjustments. Thus, a balance between the flexibility of multi-stage models and the simplicity of two-stage approaches is essential for effective long-term investment planning.

Adaptive two-stage programming offers such a balance. This framework allows initial decisions to be revised once, thereby accommodating deviations from initial forecasts while maintaining practical implementability \cite{ATSBestie}. For example, \cite{ATSShi} applies this method to the allocation of mobile energy resources during earthquakes. Nonetheless, this work focuses exclusively on a single hardening strategy and a single hazard type. The need to explore the complexities and advantages of deploying multiple hardening strategies in response to both above-ground and underground hazards remains an important open research problem.

\subsection{Our contribution}
In this work, we address the risks posed to power lines by multiple natural hazards, specifically earthquakes and high winds, and explore the use of multiple grid hardening strategies to enhance the resilience of distribution networks. Our approach integrates undergrounding (UG)  power lines and vegetation management (VM), two widely adopted yet fundamentally different strategies that mitigate distinct forms of vulnerability. UG is a capital-intensive solution with higher repair costs compared to overhead lines, as locating and repairing faults in underground infrastructure typically requires more time and resources. Consequently, failure to account for potential damage from underground hazards, such as earthquakes, can lead to significantly elevated lifecycle costs.

At the same time, combining multiple hardening strategies can create synergies that address complementary risk profiles. For instance, because underground cables are shielded from falling trees and encroaching vegetation, UG can eliminate the need for VM along these segments, generating substantial long-term cost savings. Conversely, proactive VM can significantly reduce the likelihood of outages in overhead lines, which remain susceptible to both vegetation-related incidents and seismic activity. This interplay between UG and VM not only strengthens the grid’s physical resilience but also optimizes operational and maintenance expenditures over time.

To further enhance planning efficiency, we adopt an adaptive two-stage optimization framework that allows for the revision of early-stage decisions prior to implementing irreversible actions such as undergrounding. By incorporating flexibility into the decision-making process, this approach ensures that hardening investments can be adjusted in response to new information and evolving risk conditions, thereby improving the robustness and cost-effectiveness of long-term resilience planning.

The main contributions of our work are summarized as follows:
 \begin{itemize}
    \item We present a comprehensive model for optimal grid hardening that incorporates costs associated with multiple hazards, specifically high winds and earthquakes, while integrating both long-term undergrounding and short-term vegetation management strategies. By considering a range of hazards, our model facilitates more informed decision-making regarding the implementation of expensive hardening measures. Furthermore, the integration of multiple complementary strategies enhances overall system resilience and cost-effectiveness.
    
    \item To support flexible and robust investment decisions under uncertainty, we introduce an adaptive two-stage stochastic optimization framework for grid hardening. This framework allows for a single revision of initial decisions in response to realized future scenarios, enabling the model to effectively adapt to evolving climate risks and improving long-term infrastructure planning.
\end{itemize}
    
The remainder of this paper is organized as follows: Section \ref{ATS} introduces the concept of adaptive stochastic programming. Section \ref{PF} presents the formulation of the optimal grid hardening problem. Section \ref{results} provides numerical results. Finally, Section \ref{conclusion} offers concluding remarks.

\section{Adaptive stochastic programming}
\label{ATS}
Optimizing decision-making under uncertainty is inherently complex, especially in sequential and multi-period problem settings. In the following subsections, we first review the general structure of a scenario tree and introduce the concept of adaptive two-stage programming for modeling stochastic processes. We then present tree pruning techniques designed to improve computational tractability.

\subsection{Scenario tree construction}
\begin{figure}[tb!]
	\centering
	\includegraphics[scale=0.26]{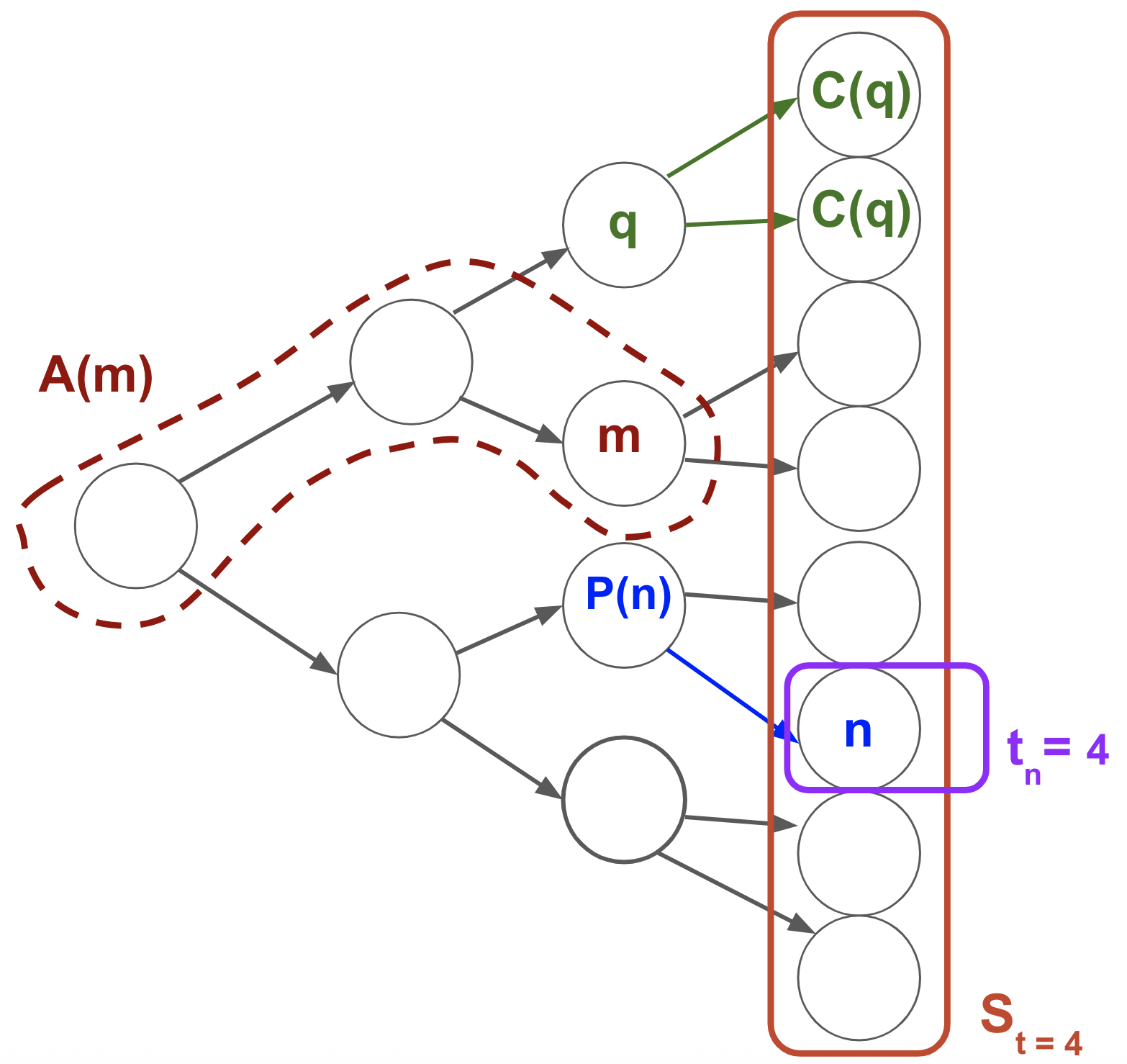}
	\caption{Scenario tree annotated with notation: parent node $P(n)$, the set of children nodes $C(n)$, ancestor nodes $A(n)$, set of nodes belong to time $t$ as $S_{t}$, and the time belongs to node $n$ as $t_{n}$.}
	\label{fig:st}
\end{figure}

A scenario tree is a foundational tool for modeling uncertainties in sequential decision-making processes, where each node represents a possible realization of the underlying uncertainty \cite{shapiro2003}. We denote the scenario tree as $\mathcal{T}$ and use it to capture the stochastic nature of the problem across $T$ time periods. A basic scenario tree is illustrated in Fig. \ref{fig:st}, with nodes labeled as $n$. The set of nodes at time period $t$ is denoted by $S_t$, for $1 \leq t \leq T$, and the time period associated with node $n$ is indicated by $t_n$.

Each unique path from the root to a leaf node constitutes a scenario, and thus the total number of scenarios equals the number of nodes in $S_T$. Every node $n$ (except the root) has a parent, denoted by $P(n)$, and a set of children nodes, denoted by $C(n)$. The set of all nodes along the path from the root to a given node $n$ is represented by $A(n)$. For any terminal node $n \in S_T$, $A(n)$ corresponds to all nodes in a complete scenario. The subtree rooted at node $n$ and extending to period $T$ is denoted by $\mathcal{T}(n)$. Each node $n$ is assigned a probability $p_n$, with the condition that $\sum_{n \in S_t} p_n = 1$ for all $1 \leq t \leq T$.

\subsection{General structure of adaptive two-stage programming}
Two-stage (TS) and multi-stage stochastic programming are classical frameworks for modeling uncertainty, where each stage corresponds to a decision point along the time horizon. In the TS approach, decisions are split into two sets: first-stage decisions are made at the start of the planning period, before any uncertainty is realized, while second-stage decisions are made in response to realized outcomes.

Over extended planning horizons, fixed first-stage decisions can become rigid and suboptimal, as they ignore evolving information. Adaptive two-stage (ATS) programming addresses this limitation by allowing first-stage decisions to be revised once at a designated revision time. Before the revision time, decisions follow the initial plan; at the revision time, the model incorporates observed uncertainty to update remaining first-stage decisions.

\begin{figure*}[tb!]
	\centering
	\includegraphics[scale=0.30]{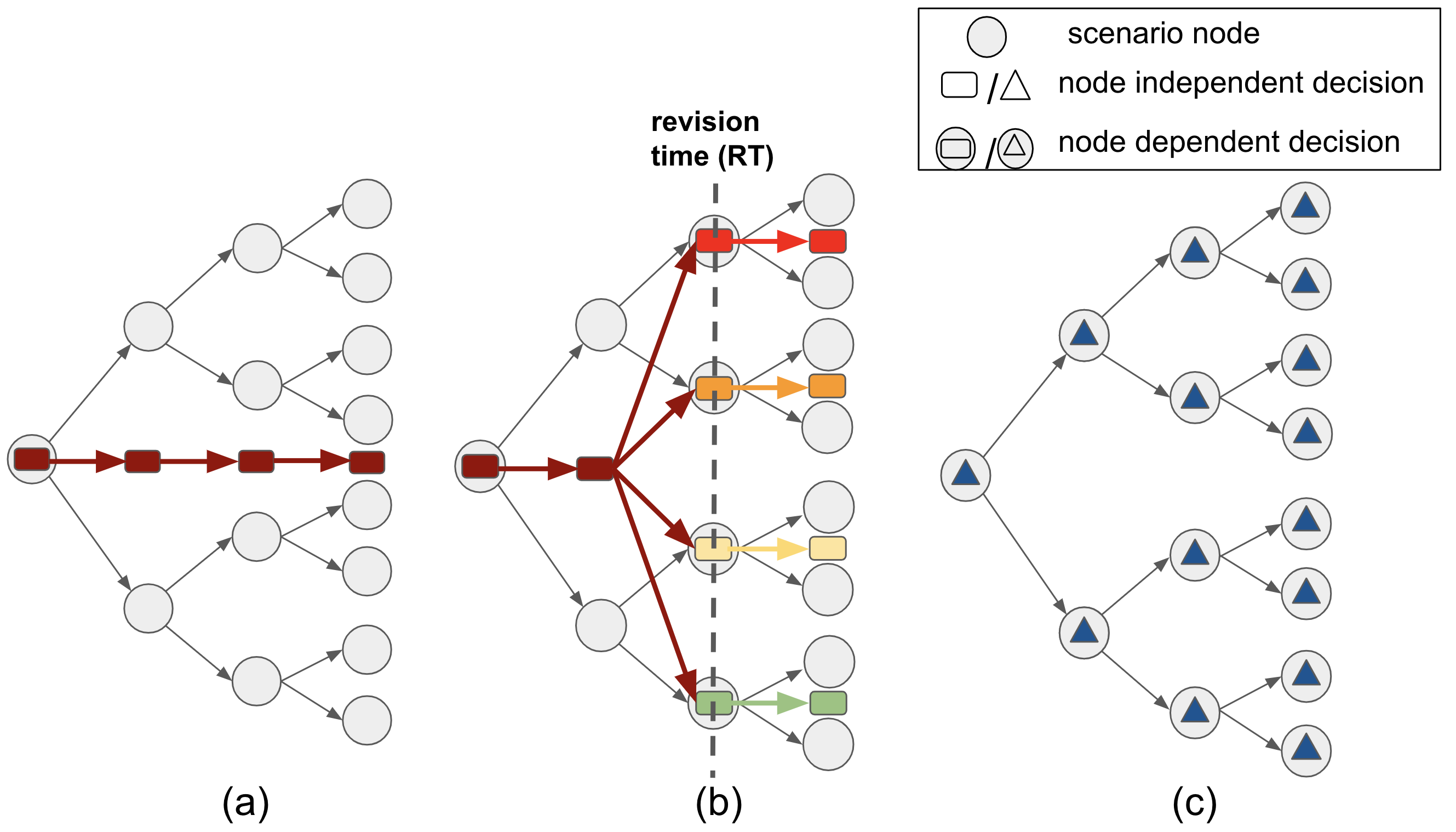}
	\caption{Decision structure of first-stage variable $(\sqsubset \! \sqsupset)$ (a) two-stage; (b) adaptive two-stage; and (c) decision structure of second-stage variable $(\triangle)$ for both two-stage and adaptive two-stage.}
	\label{fig:dt}
\end{figure*}

Figs.~\ref{fig:dt}(a) and \ref{fig:dt}(b) highlight the difference between the standard two-stage and ATS frameworks. In the standard model, decisions are fixed for all time steps. In contrast, ATS permits branching at the revision time, allowing scenario-specific updates. After the revision time, decision-making proceeds as in the standard model but along each scenario path.
Importantly, the revision mechanism applies only to first-stage decisions. In both models, second-stage decisions remain scenario-dependent, as shown in Fig. \ref{fig:dt}(c).

\subsection{Tree pruning}
\begin{figure*}[tb!]
	\centering
	\includegraphics[scale=0.30]{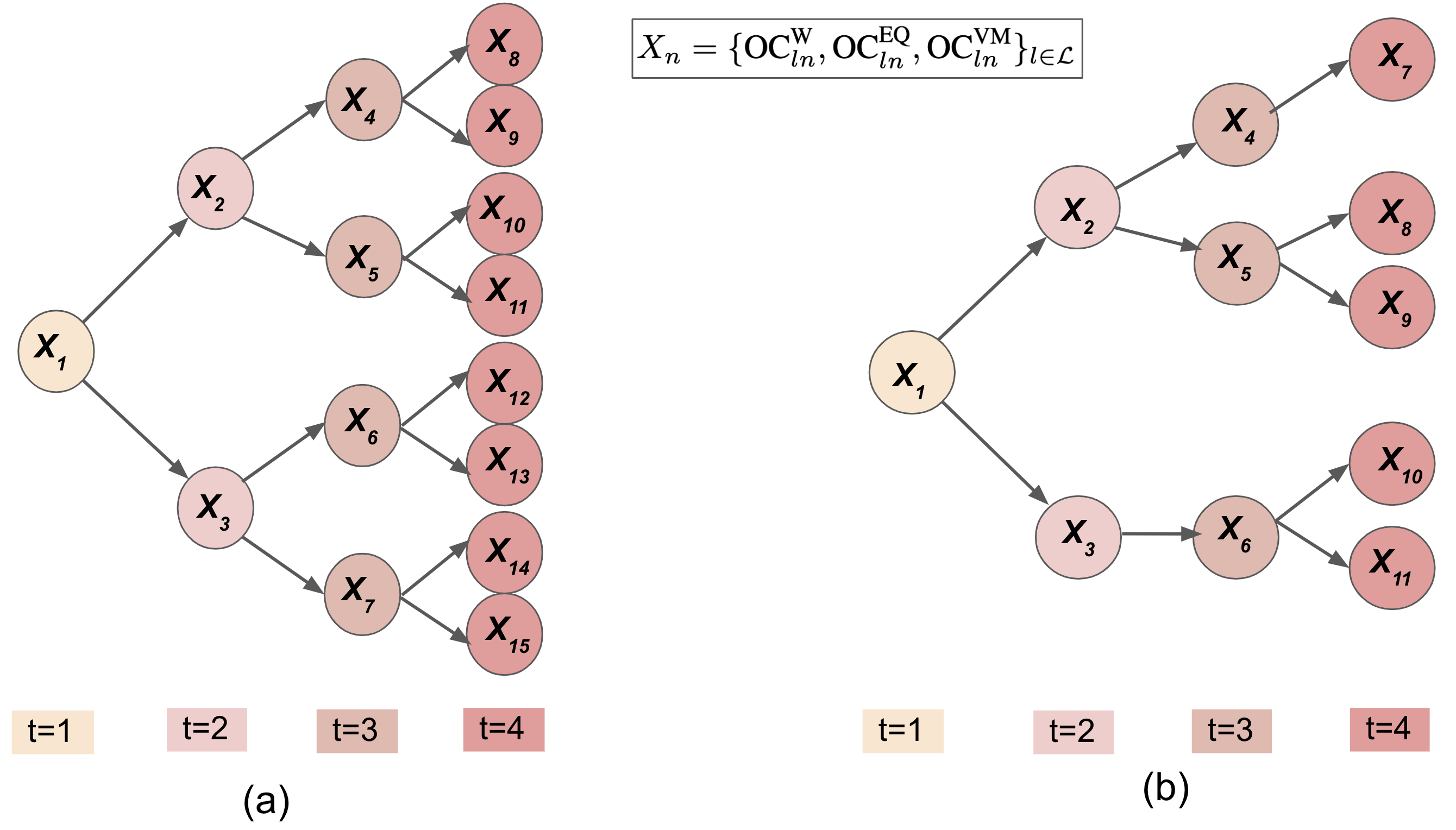}
	\caption{For $T=4$, (a) full scenario tree and (b) pruned scenario tree.}
	\label{fig:tree type}
\end{figure*}

As the planning horizon extends, the scenario tree grows exponentially, making it computationally intensive to model the associated stochastic process. To mitigate this complexity, scenario reduction techniques are used to prune the tree by eliminating less critical nodes, thereby enabling a smaller set of representative scenarios to approximate the original probability distribution \cite{dupavcova2000scenarios}. Fig. \ref{fig:tree type} provides an illustration of the full and pruned scenario trees. The uncertain parameters associated with each node $n$ are denoted by $X_n$.

Classical pruning methods include forward selection and backward reduction \cite{FSandBR}. In forward selection, the algorithm begins with an empty tree and iteratively adds scenarios that yield the greatest improvement in approximation quality. Conversely, backward reduction starts with the full scenario tree and iteratively removes nodes whose exclusion results in the smallest loss of fidelity. Both approaches continue until a predefined number of scenarios is reached. In this work, we adopt the backward reduction method to prune the original scenario tree and ensure that the simulation remains computationally tractable \cite{scenred}. 

\section{Problem formulation}\label{PF}
We formulate the optimal grid hardening problem as an adaptive two-stage stochastic program on a radial distribution system with distributed energy resources and a connection to the upstream transmission network. The objective is to balance long-term electricity shortfall costs with investments in two hardening strategies: undergrounding overhead lines and vegetation management. 
 
\subsection{Load shedding cost of each outaged line}\label{sec:LS cost}

We consider single-line outages caused by either extreme wind (W), earthquakes (EQ), or vegetation (VM) related events, denoted by the index $e$. To compute the daily load shedding cost $\text{C}_{l}^{\text{LS,e}}$ for each distribution line $l\in \mathcal{L}$,  during an event $e$, we formulate the following optimization problem where $\mathcal{H}\triangleq \{1,2,\dots,24\}$, $P^{\text{L,e}}_{i,h}$ is the active load demand at bus $i$, during event $e$ and time $h$, $\text{C}^\text{ENS}$ is the cost of energy not served.
\begin{subequations}
\begin{align}
\text{C}_{l}^{\text{LS,e}} := \min \quad &\sum_{h \in  \mathcal{H}} \sum_{i} P_{i,h}^{\text{L,e}} \text{C}^{\text{ENS}}
\label{Eqtn: LS calculation}\\
\text{s.t.}\quad
&P_{i,h} = \sum_{k:i\,\to\,k} P_{ik,h} - \sum_{j:j\,\to\,i} \left(P_{ji,h} - r_{l}\ell_{l,h} \right),\, \forall h \in  \mathcal{H}
\label{pf_1}\\
&Q_{i,h} = \sum_{k:i\,\to\,k} Q_{ik,h} - \sum_{j:j\,\to\,i} \left(Q_{ji,h} - x_{l}\ell_{l,h} \right),\, \forall h \in  \mathcal{H}
\label{pf_2}\\
&v_{i,h} = v_{j,h} - 2 \left(r_{l}P_{l,h} + x_{l}Q_{l,h}\right) + (r_{l}^{2}+x_{l}^{2})\ell_{l,h},\, \forall h \in  \mathcal{H}
\label{pf_3}\\
&P_{l,h}^{2} + Q_{l,h}^{2}  \le \ell_{l,h} v_{j,h},\, \forall h \in  \mathcal{H} \label{pf_4}\\
&\underline{v}\leq v_{i,h}\leq \bar{v}, \quad v_{i,h}^{\text{ref}}=1, \quad 0\le \ell_{l,h} \times u_{l}\leq \bar{\ell}_{l},\, \forall h \in  \mathcal{H}
\label{voltage_limit}\\
&0 \leq P_{i,h} \leq \bar{P}^{\text{MT}}_{i}, \quad \forall i \in \mathcal{N}^{\text{MT}},\, \forall h \in  \mathcal{H}
\label{MT limit} \\
&P^{\text{MT,rd}}_{i} \leq P_{i,h+1} - P_{i,h}\leq P^{\text{MT,ru}}_{i},\, \quad\forall i \in \mathcal{N}^{\text{MT}},\,\forall h \in  \mathcal{H} \setminus \{24\}.
\label{ramping up-down limit}
\end{align}
\label{prob:LScost}
\end{subequations}

To minimize the total daily load shedding cost across all lines, as defined in \eqref{Eqtn: LS calculation}, constraints \eqref{pf_1}–\eqref{pf_4} enforce power flow consistency throughout the network based on the \emph{DistFlow} model. Constraints \eqref{voltage_limit} establishes lower and upper bounds on nodal voltages and branch currents to ensure safe system operation. The binary indicator $u_{l}$ is used to represent the de-energized state of line $l$ (i.e., $u_{l} = 0$), enabling evaluation of the impact of a specific line outage on the rest of the network. Finally, constraints \eqref{MT limit}–\eqref{ramping up-down limit} impose limits on power generation and ramping capabilities of microturbine units. Note that the above optimization problem is solved offline, and its results are used as input parameters in the grid hardening formulation presented in the subsequent sections.

\subsection{Objective function}
 We define $\bm{\Xi} \triangleq \{\xi_{n}^{\text{e}}, N_{n}^{\text{e}}\}_{n \in \mathcal{T}, e\in \{{\text{W},\text{EQ},\text{VM}}\}} $as the set of all random variables, where the subscript $n$ denotes the index of a scenario node. For each line $l$, node $n$ and a given event type $e$, the outage cost $\text{OC}^{\text{e}}_{ln}$ is calculated as the product of the daily load shedding cost $\text{C}_{l}^{\text{LS,e}}$, the failure probability $\varrho^{\text{e}}_{ln}$, the expected outage duration $\xi^{\text{e}}_{n}$, and the predicted number of events $N^{\text{e}}_{n}$ in a year. These costs are defined in \eqref{c_oc}, and are denoted as $X_n$ in the pruned scenario tree (see Fig. \ref{fig:tree type}). The corresponding total costs, which include both outage $\text{OC}$ and repair costs $\text{RC}$ over the planning horizon, are given by $\text{TC}^{\text{e}}$ in \eqref{Eqtn:tc} where $p_{n}$ is the probability of node $n$. To guide investment prioritization, we introduce two scaling factors. $\delta_{ln}^{\text{VM}}$ reflects the relative importance of vegetation-related risks for line $l$ at node $n$, while $\delta_{ln}^{\text{UG}}$ captures the ratio of projected earthquake to wind-related costs. Finally, \eqref{Eqtn: discount rate} defines the nominal discount factor $\gamma$ based on the inflation rate $f$ and the discount rate $d$.
\begin{align}
\text{OC}_{ln}^{\text{e}}(\bm{\Xi}) &= \text{C}_{l}^{\text{LS,e}} \times \varrho^{\text{e}}_{ln}\times \xi^{\text{e}}_{n}  \times N^{\text{e}}_{n},\, \quad \forall e \in  \{\text{W},\text{EQ},\text{VM}\}
\label{c_oc}\\
\text{C}^{\text{e}}_{ln}(\bm{\Xi}) &= \text{OC}_{ln}^{\text{e}} +\text{RC}^{\text{e}}_{ln},\, \quad \forall e \in  \{\text{W},\text{EQ},\text{VM}\}
\label{Eqtn:OH cost}\\
\text{TC}^{\text{e}}_{ln}(\bm{\Xi})&= \sum_{m\in \mathcal{T}(n)} \frac {p_{m}}{p_{n}} \times \text{C}_{lm}^{\text{e}} ,\, \quad \forall e \in  \{\text{W},\text{EQ},\text{VM}\}
\label{Eqtn:tc}\\
\delta_{ln}^{\text{VM}}(\bm{\Xi}) &= \frac{\sum_{m\in \mathcal{T}(n)} p_{m} \text{C}_{lm}^{\text{VM}}}{\text{max}_{l \in \mathcal{L}} \Bigg(\sum_{m\in \mathcal{T}(n)}p_{m} \text{C}_{lm}^{\text{VM}}\Bigg)},\, \quad \delta_{ln}^{\text{UG}} (\bm{\Xi}) = \frac{\sum_{m\in \mathcal{T}(n)}p_{m} \text{C}_{lm}^{\text{EQ}}} {\sum_{m\in \mathcal{T}(n)} p_{m} \text{C}_{lm}^{\text{W}}},\,
\label{delta vm}\\
\gamma &=\frac{1+f}{1+d}.
\label{Eqtn: discount rate}
\end{align}

The planning objective is to minimize the total cost defined in \eqref{Eqtn:obj}, which accounts for undergrounding (UG) and vegetation management (VM) grid hardening strategies. UG is modeled as a long-term investment with a service life of 30 years, while VM is treated as a short-term measure implemented annually. The binary variable $\alpha_{ln}$ represents a first-stage decision that indicates whether line $l$ at scenario node $n$ is selected for undergrounding. The continuous second-stage variable $\beta_{ln}$ specifies the proportion of vegetation to be managed around line $l$ at node $n$. The binary variable $r_{l,t}$ identifies the revision time $t$ at which the initial undergrounding decision $\alpha_{ln}$ for line $l$ may be updated.

Equations \eqref{eqtn:UG} and \eqref{eqtn:VM} represent the net costs of UG and VM, respectively. In \eqref{eqtn:UG}, $\text{IC}^{\text{UG}}_{ln}$ is the direct per unit undergrounding cost for line $l$ at node $n$. Earthquake-related outage and repair costs are still incurred after undergrounding, while wind- and vegetation-related costs are eliminated. The final term accounts for wind-related costs before undergrounding is implemented. In \eqref{eqtn:VM},  the first term stands for the VM implementation cost while the last term captures the residual outage costs due to incomplete vegetation management.
\begin{equation} 
{\text{minimize}}\quad  Q^{\text{UG}}\left(\alpha_{ln}\right) + Q^{\text{VM}}\left(\beta_{ln}\right)
\label{Eqtn:obj}
\end{equation}
where 
\begin{subequations}
\begin{align}
&Q^{\text{UG}}= \sum_{n\in \mathcal{T}} p_{n} \times \gamma^{t_{n}-1} \sum_{l \in \mathcal{L}}  \alpha_{ln} \times \left( \mu_{l} \times
\text{IC}^{ \text{UG}}_{ln}+\text{TC}^{\text{EQ}}_{ln} - \text{TC}^{\text{W}}_{ln} - \text{TC}^{\text{VM}}_{ln} \right)  \notag  \\ & \hspace{7.5cm}+ \left(1- \sum_{m\in A(n)} \alpha_{lm}\right)  \text{C}_{ln}^{\text{W}}  
\label{eqtn:UG}\\
&Q^{\text{VM}}= \sum_{n\in \mathcal{T}} p_{n} \times \gamma^{t_{n}-1}    \sum_{l \in \mathcal{L}} \beta_{ln} \times \mu_{l}\times\text{IC}^{\text{VM}}_{ln}  + \left( 1- \beta_{ln} -\sum_{m\in A(n)} \alpha_{lm} \right)\text{C}_{ln}^{\text{VM}}
\label{eqtn:VM}
\end{align}
\end{subequations}
\vspace{-2em}

\subsection{Constraints}
The grid hardening constraints are given as follows:
\begin{subequations}
\begin{align}
& \sum_{l \in \mathcal{L}} \alpha_{ln} \leq N^\text{UG}_{n} ,  \forall n \in \mathcal{T} 
\label{Eqtn:maxm number of lines to be UGed in each n} \\
&\sum_{n \in \mathcal{T}} p_{n} \sum_{l \in \mathcal{L}} \alpha_{ln} \times \gamma^{t_{n}-1} \times \mu_{l} \times \text{IC}^{{\text{UG}}}_{ln} \leq B^{\text{UG}}
\label{Eqtn: total budget for UG }\\
& \sum_{m \in A(n)} \alpha_{lm} \leq 1, \quad \forall l \in  \mathcal{L}, \forall n \in S_{T}
\label{Eqtn: non-changeable status of a UG line}\\ 
&\sum_{m\in A(n)} \alpha_{lm} + \beta_{ln} \leq 1, \quad \forall l \in \mathcal{L},\forall n \in  \mathcal{T}
\label {Eqtn: Coupling constraint between UG and VM}\\
&0 \leq \beta_{ln} \leq \delta_{ln}^{\text{VM}} \times \rho_{ln}, \quad \forall l \in \mathcal{L},\forall n \in  \mathcal{T}
\label{Eqtn:bb}\\ 
 & \sum_{n \in \mathcal{T}} p_{n}  \sum_{l \in \mathcal{L}} \beta_{ln} \times \gamma^{t_{n}-1} \times \mu_{l} \times \text{IC}^{{\text{VM}}}_{ln} \leq  B^{\text{VM}}
\label{Eqtn: total budget for VM }\\
& \sum_{t=1}^{T} r_{lt}=1,\quad \forall l \in  \mathcal{L}
\label{Eqtn: revision decision}\\
&\alpha_{ln} + \left(1-\sum_{t^{\prime}=t+1}^{T} r_{lt^{\prime}}\right) \ge \alpha_{lm}\geq \alpha_{ln} - \left(1-\sum_{t^{\prime}=t+1}^{T} r_{lt^{\prime}}\right),\, \notag  \\ & \hspace{5.5cm} \forall m,n \in S_{t}, t \in {\tilde{\mathcal{T}}} \setminus \{T\},  \forall l \in \mathcal{L}
\label{Eqtn: 1st inequality for r}\\
&\alpha_{ln} + (1- r_{lt}) \ge \alpha_{lm}\geq \alpha_{ln} - (1- r_{lt}),\, \notag  \\ & \hspace{5.5cm}  \forall m,n \in S_{t^{\prime}} \cap \mathcal{T}(q), q \in S_{t}, t^{\prime}\geq t, t\in {\tilde{\mathcal{T}}}, \forall l \in \mathcal{L}
\label{Eqtn: 3rd inequality for r}\\
&\alpha_{ln}, r_{lt}\in \{0,1\}, \quad \forall l \in \mathcal{L}, \forall t \in {\tilde{\mathcal{T}}}.
\label{Eqtn: r as a binary variable}
\end{align}
\end{subequations}

Constraint \eqref{Eqtn:maxm number of lines to be UGed in each n} limits the number of lines that can be undergrounded at each scenario node, while \eqref{Eqtn: total budget for UG } imposes a system-wide UG budget constraint $B^{\text{UG}}$. In \eqref{Eqtn: non-changeable status of a UG line}, $T$ denotes the last time period in the planning horizon and this constraint enforces that once a line is undergrounded, the decision cannot be reversed. Constraint \eqref{Eqtn: Coupling constraint between UG and VM} links the first- and second-stage decisions, permitting vegetation management only for overhead lines. In \eqref{Eqtn:bb}, $\rho_{ln}$ is the observed vegetation percentage. This constraint determines the required VM level along line $l$ at node $n$, scaled by the line’s criticality $\delta^{\text{VM}}$ as defined in \eqref{delta vm}. The total VM budget $B^{\text{VM}}$ is used in constraint \eqref{Eqtn: total budget for VM }.

Equations \eqref{Eqtn: revision decision} to \eqref{Eqtn: 3rd inequality for r} define the logic governing the revision decision variable $r_{l,t}$. Constraint \eqref{Eqtn: revision decision} ensures that each line is subject to at most one revision throughout the planning horizon. Constraints \eqref{Eqtn: 1st inequality for r} and \eqref{Eqtn: 3rd inequality for r} describe the structure of first-stage decisions before and after the revision point, respectively. ${\tilde{\mathcal{T}}}$ denotes the set of years in the planning horizon. As illustrated in Fig. \ref{fig:dt}(b), before the revision occurs (i.e., when $r_{l,t} = 1$ has not yet been triggered), a single first-stage decision $\alpha_{l}$ is applied uniformly to all scenario nodes at time $t$, denoted as the set $S_{t}$. Once the revision time is reached, with scenario-specific outcomes and updated forecasts of outage costs available, the value of $\alpha_{l}$ can differ across scenarios to reflect observed conditions. These revised decisions then remain fixed through the end of the planning period $T$. Finally, constraint \eqref{Eqtn: r as a binary variable} enforces the binary nature of both the undergrounding decision variable $\alpha$ and the revision indicator $r$.

It is worth noting that the grid hardening problem formulation is topology agnostic and does not depend on whether the network is radial or meshed. However, to compute load shedding costs, optimization problem~\eqref{prob:LScost} uses the DistFlow model, which assumes a radial network structure commonly found in distribution systems. For weakly meshed networks, one can adopt Shor relaxations~\cite{Molzahn-Hiskens-2019} or generalized DistFlow models~\cite{Masoud-Low-2013} that approximate behavior in such settings. Alternatively, for general meshed grids, full AC power flow formulations can be employed to ensure physical accuracy.

\section{Simulation results}
\label{results}
This section presents simulation results that demonstrate the effectiveness of the proposed framework.

\subsection{Simulation setup}
  \begin{figure}
  \centering
  \captionsetup[subfigure]{justification=centering}
  \begin{subfigure}{0.45\columnwidth}
  \includegraphics[scale=0.27]{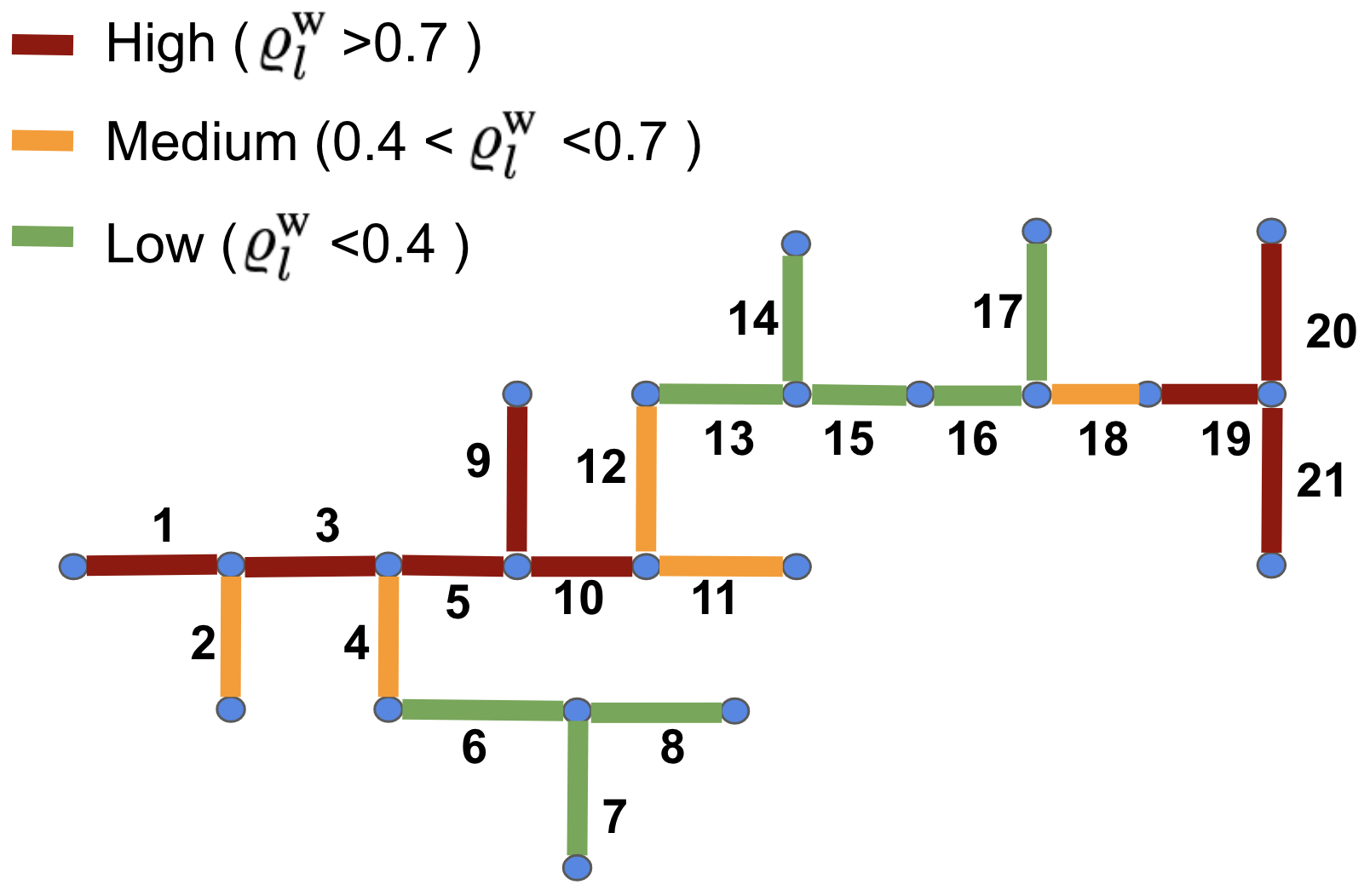}
  \caption{}
  \label{fig:risk wind}
  \end{subfigure}
  \hfill
  \begin{subfigure}{0.45\columnwidth}
  \includegraphics[scale=0.27]{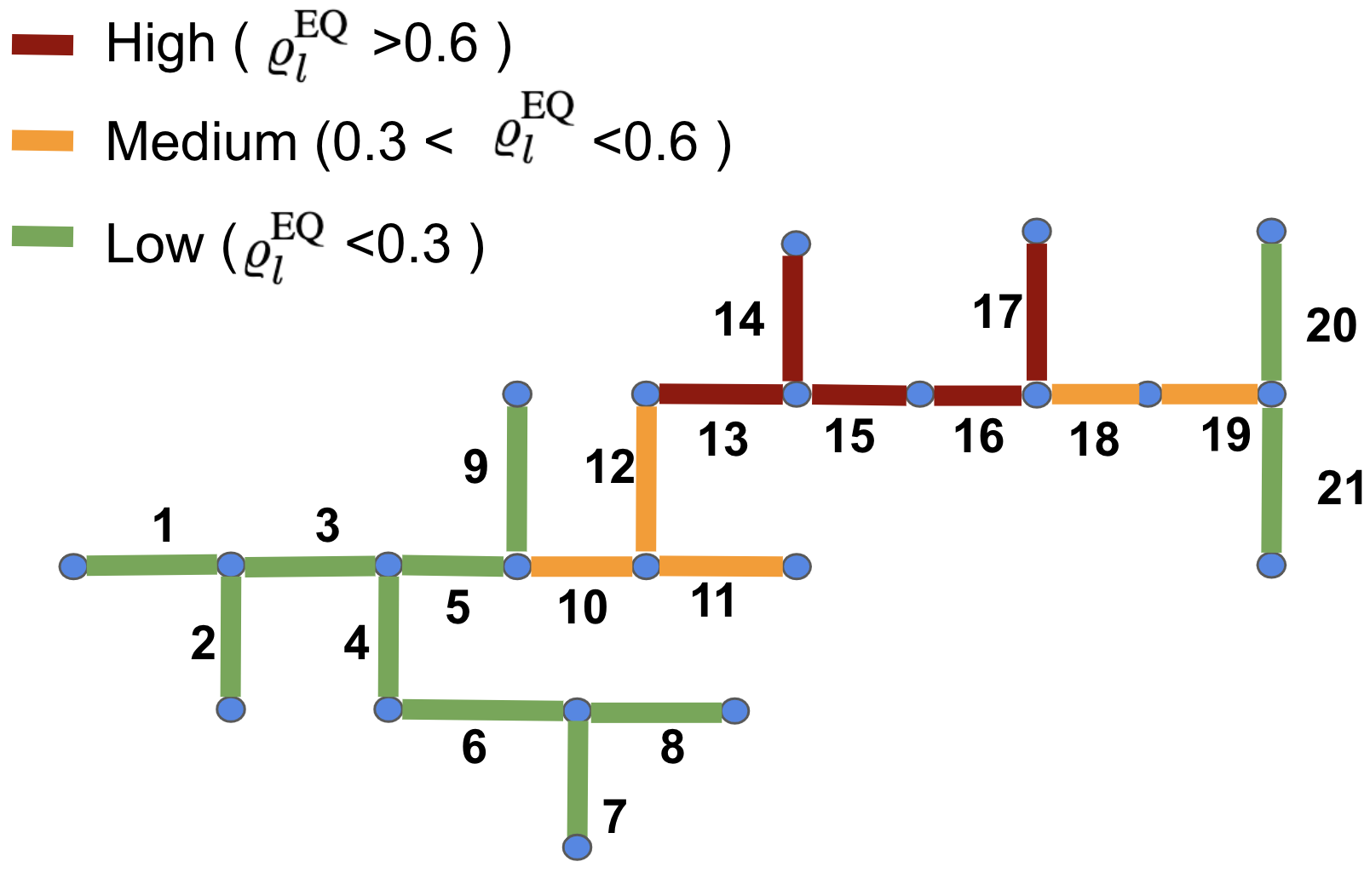}
  \caption{} 
  \label{fig:risk eq}
  \end{subfigure} 
  \caption{Risk level for power lines facing (a) high wind (b) earthquake.}
  \end{figure}

\renewcommand{\arraystretch}{0.5}
\begin{table}[tb!]
 \centering
 \caption{Parameters of the grid hardening problem.}
 \small
 \begin{tabular}{c|c}
    \hline
    \textbf{Parameter} & \qquad \textbf{Value}\\  
    \hline
    $\text{C}^{\text{ENS}}$ & \$10,000/MWh \\
    \hline
    $\text{IC}^{\text{UG}}_{n}, \text{IC}^{\text{VM}}_{n}$ & \$3.4M/mile, \$2,275/mile \\
    \hline
    $T$ & 30 years\\
    \hline
    $f, d$ & 0.03, 0.02\\
    \hline
    $B^{\text{VM}}$ & \$430K\\
    \hline
\end{tabular}
\label{Tab:parameters}
\end{table}

All simulations are conducted in MATLAB on a personal computer equipped with an Intel Core i7 processor 3.6 GHz and 32 GB of RAM. Gurobi v12.0.0 is employed as the optimization solver, interfaced through YALMIP. A modified IEEE 22-bus distribution system is used to evaluate the proposed model. 

The outage duration parameters $\xi^{\text{W/EQ/VM}}$ are modeled using lognormal distributions, with mean values of 25.5 hours for high wind events, 20 days for earthquakes, and 3 hours for outages caused by fallen trees \cite{distribution_outage_duration}. The number of events $N^{\text{W/EQ/VM}}$ is treated as a discrete random variable following a Poisson distribution. Data for modeling $N^{\text{W}}$ is obtained from \cite{number_projection}. For simplicity, the failure probabilities of different lines, $\varrho^{\text{W/EQ/VM}}$, are assigned as fixed values. Note that our framework can readily adopt failure probability estimates derived from empirical data, physics-based models or probabilistic hazard analyses. We consider a typical day load profile $P^{\text{L}}_{i,h}$, irrespective of any particular type of hazard $e$, to calculate the daily load shedding cost $C_{l}^{\text{LS}}$. The purpose of this calculation is to approximate the
financial impact of service interruptions, irrespective of the specific cause of the
outage. Moreover, our model is flexible to accommodate distributed energy resources by integrating the corresponding operational constraints into problem~\eqref{prob:LScost}, without affecting the overall computation time. Additionally, all distribution lines are assumed to have equal length. The reported optimal objective values $v^{\text{ATS}}$ and $v^{\text{TS}}$ are in million dollars. The complete set of simulation parameters is given in Table \ref{Tab:parameters}.

\subsection{Optimal line hardening decision}
\begin{figure}[tb]
	\centering
	\includegraphics[scale=0.70]{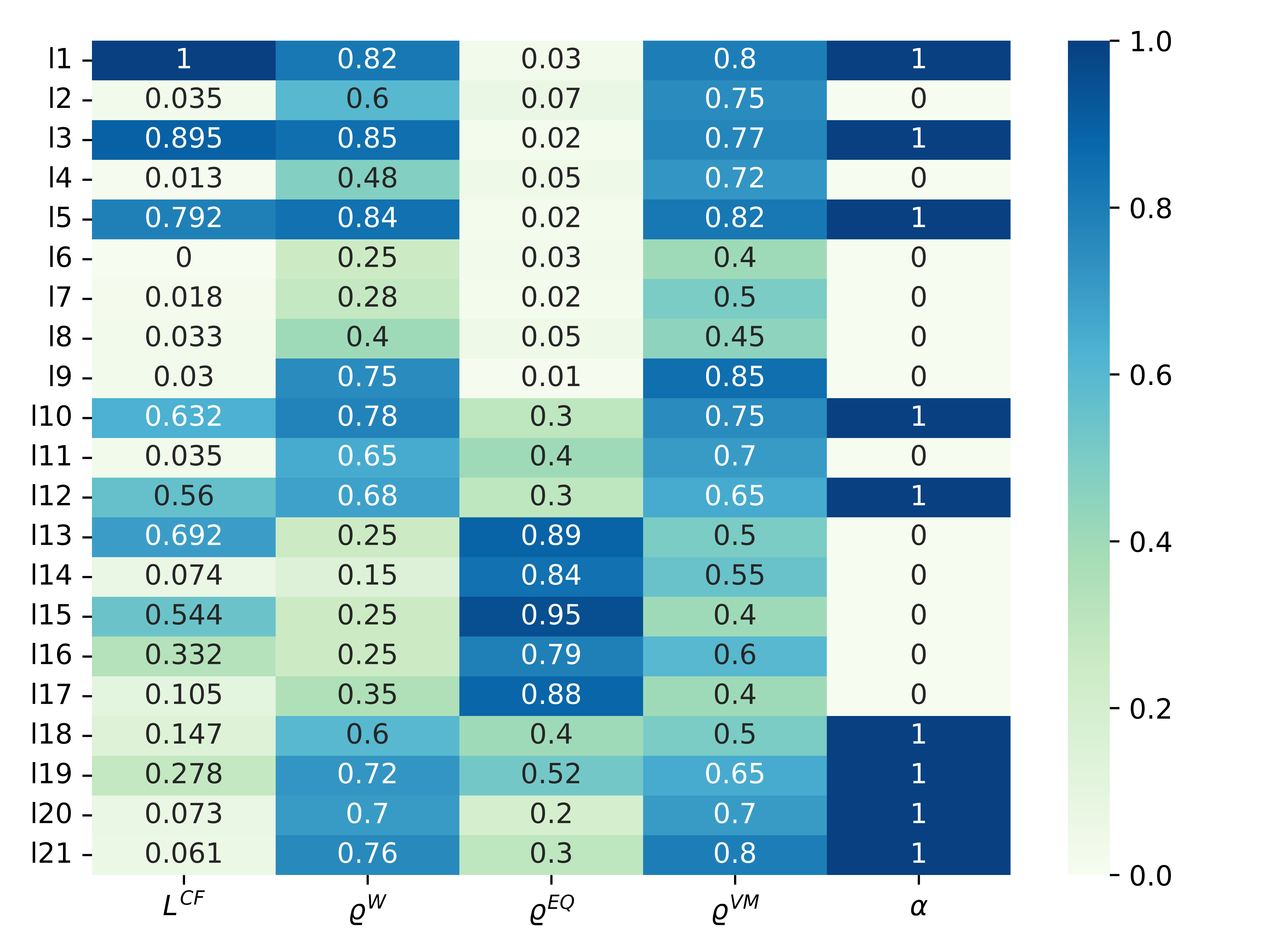}
	\caption{Line undergrounding decision $(\alpha)$ with respect to relative load shedding cost $(\text{L}^\text{CF})$, line outage probability for high wind $(\varrho^{\text{W}})$, earthquake $(\varrho^{\text{EQ}})$ and fallen trees $(\varrho^{\text{VM}})$.}
	\label{fig:heatmap}
\end{figure}

Based on the load shedding costs, we define the line criticality factor $\text{L}^{\text{CF}}$ for each line as 
$\text{L}^{\text{CF}}_l = \frac{\text{C}_{l}^{\text{LS}}}{\text{max}_{l \in \mathcal{L}} \{\text{C}_{l}^{\text{LS}}\}}$.
Figs.~\ref{fig:risk wind} and \ref{fig:risk eq} illustrate the categorical risk levels of each line under wind and earthquake conditions, respectively. 
Fig. \ref{fig:heatmap} presents the optimal hardening decisions across all lines, based on their relative load shedding cost $\text{L}^{\text{CF}}_l$ and associated disaster risk levels. Lines with high $\text{L}^{\text{CF}}_l$ and low earthquake-to-wind outage cost ratios $\delta^{\text{UG}}_l$ are prioritized for undergrounding.

For instance, with a \$35 million undergrounding budget, lines 1, 3, 5, 10, and 12 are selected due to their high load shedding impact, wind vulnerability, and relatively low earthquake risk. In contrast, lines 13 and 15 are excluded despite their high $\text{L}^{\text{CF}}_l$ values, as they are more prone to seismic damage. Lines 14, 16, and 17 are also omitted due to high $\varrho^\text{EQ}$. As the budget increases, additional wind-vulnerable lines, such as 18 to 21, are incorporated. This expanded budget allows for the inclusion of lower-priority lines initially excluded from the plan.

Table \ref{Tab:bug vs lines} illustrates the effect of increasing the undergrounding budget on the optimal selection of hardened lines. For instance, with a budget of \$27 million allocated to undergrounding, seven lines are selected. As the budget increases, additional lines 20, 21, 9, and 2 are sequentially added to the hardening plan based on their criticality and outage probabilities. Expanding the undergrounding budget not only increases the number of protected lines but also leads to a significant reduction in expected outage and repair costs, as reflected by the declining objective values $v^{\text{ATS}}$. Moreover, as more lines are placed underground, the need for vegetation management along those segments is eliminated, resulting in lower VM costs and contributing to improved overall system efficiency.

\renewcommand{\arraystretch}{0.6}
\begin{table*} [tb]
 \caption{Optimal selection of lines with varying budget. The bold numbers indicate the indices of newly added lines as the budget increases.}
 \label{Tab:bug vs lines}
 \centering
 \small
 \begin{tabular}{c|c|c|c}
    \hline
     $B^{\text{UG}}$& 
     \text{Indices of selected lines for hardening} & $v^{\text{ATS}}$ & \text{VM cost} ($\$\times 10^{3}$)\\ 
    \hline
    27 & 1,3,5,10,12,18,19  & 309.22 & 94.93\\
    \hline
    31 & 1,3,5,10,12,18,19,\textbf{20}  & 258.59 & 91.57 \\
    \hline
    35 & 1,3,5,10,12,18,19,20,\textbf{21} & 219.78 & 89.55\\
    \hline
    39 & 1,3,5,\textbf{9},10,12,18,19,20,21 & 188.06 & 88.12\\
    \hline
    43 & 1,\textbf{2},3,5,9,10,12,18,19,20,21 & 159.67 & 86.60\\
    \hline
\end{tabular}
\end{table*}

\subsection{Comparison of ATS with classic two-stage optimization model}
Table \ref{Tab:ts vs ats} summarizes the benefits of ATS compared to standard TS programming across three undergrounding budget levels and two scenario tree sizes. ATS reduces to TS when all revision indicators are fixed as $r_{l,t} = 1$. The relative improvement, defined as $G_{\text{rel}} = \frac{v^{\text{TS}} - v^{\text{ATS}}}{v^{\text{TS}}} \times 100\%$, measures the percentage cost savings achieved by ATS over TS. The results indicate that the advantage of ATS becomes more significant with higher budgets and larger scenario trees. In these cases, the decision space is broader and the uncertainty more pronounced, amplifying the value of flexibility in revising initial decisions. Although ATS incurs a higher computational burden, the additional effort is justified given the long-term nature of infrastructure planning and the substantial financial benefits it delivers. The increase in computation time remains acceptable in light of the improved planning outcomes.

Although the ATS model shows superior performance under the aforementioned conditions, TS may be sufficient as a less computationally intensive choice in certain contexts where post-decision adjustments become insignificant. For example, under a shorter planning horizon, the time window to observe unfolding uncertainty and revise earlier decisions becomes too narrow to allow meaningful adaptation. This diminishing value of adaptation is shown in Table \ref{Tab: ATS vs TS under different T}, where the relative gain in runtime shows little variation. Moreover, the flexibility offered by ATS provides limited marginal value over TS when uncertainty is low, such as in cases of accurately forecasted weather or when urgent investment is required due to impending hazards.

\renewcommand{\arraystretch}{0.6}
\begin{table*} [tb]
 \caption{Performance comparison between two-stage (TS) and adaptive two-stage (ATS) approach.}
 \label{Tab:ts vs ats}
 \centering
 \small
 \begin{tabular}{*8c}
   \cline{1-8}
   & & \multicolumn{3}{?c}{Objective value} & \multicolumn{3}{?c}{Computation time (seconds)}\\  
   \hline
    {$B^{\text{UG}}$} &\multicolumn{1}{|c} {Nodes}& \multicolumn{1}{?c} {$v^{\text{TS}}$} & {$v^{\text{ATS}}$} & $G_{\text{rel}}$ & \multicolumn{1}{?c} {TS} & ATS & {Increase percentage}\\ 
    \hline  
    \multirow{2}{*}{35} & \multicolumn{1}{|c}{465} & \multicolumn{1}{?c}{249.32} & 219.78 & 11.85\%  & \multicolumn{1}{?c}{95.27} & 144.46 & 34.05\%\\
    & \multicolumn{1}{|c}{1065} & \multicolumn{1}{?c}{241.56} & 208.11  & 13.85\%  & \multicolumn{1}{?c}{266.63} & 666.55 & 60.00\%\\
    \hline
    \multirow{2}{*}{39} & \multicolumn{1}{|c}{465} & \multicolumn{1}{?c}{217.17} & 188.06 & 13.40\% & \multicolumn{1}{?c}{94.11} & 184.20 & 48.91\%\\
    & \multicolumn{1}{|c}{1065} & \multicolumn{1}{?c}{207.64} & 173.99 & 16.21\% & \multicolumn{1}{?c}{260.29} & 690.64 & 62.31\%\\
    \hline
    \multirow{2}{*}{43} & \multicolumn{1}{|c} {465} & \multicolumn{1}{?c} {187.55} & 159.67 & 14.87\% &  \multicolumn{1}{?c} {96.00} & 141.00 & 31.91\%\\
    & \multicolumn{1}{|c}{1065} & \multicolumn{1}{?c} {176.42} & 144.97 & 17.83\%  & \multicolumn{1}{?c}{249.24} & 588.63 & 57.66\%\\
    \hline
    \end{tabular}
\end{table*}

\begin{table*} [tb]
 \centering
    \caption{Evaluation of TS and ATS performance under varying lengths of planning horizon, under an undergrounding budget of $B^{\text{UG}}=\$35$M.}
    \label{Tab: ATS vs TS under different T}
    \small
    \begin{tabular}{*7c}
    \cline{1-7}
    \multirow{2}{*}{$T (\text{years})$} & \multicolumn{3}{?c}{Objective value}& \multicolumn{3}{?c}{Computation time  (seconds)}\\  
    \cline{2-7}
    & \multicolumn{1}{?c} {$v^{\text{TS}}$} & $v^{\text{ATS}}$ & $G_{\text{rel}}$ & \multicolumn{1}{?c} {TS} & ATS & {Increase percentage}\\  
    \hline
    30 & \multicolumn{1}{?c}{219.78} & 249.32 & 11.84\% & \multicolumn{1}{?c}{95.27} & 144.46 & 34.05\%\\
    \hline
    20 & \multicolumn{1}{?c}{125.70} & 138.52 & 9.25\% & \multicolumn{1}{?c}{37.12} & 60.13 & 38.27\%\\
    \hline
    10 & \multicolumn{1}{?c}{105.15} & 104.83 & 0.30\% & \multicolumn{1}{?c}{10.27} & 15.3 & 32.88\%\\
    \hline
    \end{tabular}
    \end{table*}

The results discussed thus far assume a single line outage at any given time. To assess the performance of the proposed model under more complex disruption scenarios, we also consider cases involving simultaneous outages of multiple lines. In Table \ref{Tab:multiple outages}, two such cases are analyzed: one involving lines 1, 3, and 5, and another involving lines 19, 20, and 21. These concurrent outages result in higher load shedding costs due to their cascading effects on the network. Consequently, undergrounding these critical lines yields greater cost savings than in the single-line case, producing lower objective values for both TS and ATS formulations. Furthermore, the value of decision-making flexibility becomes more pronounced in the presence of multiple failures, where uncertainty, especially in the outage cost parameters $\text{OC}^{\text{W/EQ/VM}}_{ln}$, is substantially higher. As shown in Table \ref{Tab:multiple outages}, ATS offers greater relative benefits in these scenarios, underscoring its effectiveness in managing highly uncertain and high-impact grid failures.

\renewcommand{\arraystretch}{0.6} 
\begin{table} [tb]
 \caption{Performance analysis for multiple line outages, under an undergrounding budget of $B^{\text{UG}}=\$35$M.}
 \label{Tab:multiple outages}
 \centering
 \small
 \begin{tabular}{l|c|c|c}
 \cline{1-4}
     \text{Outage type} & $v^{\text{TS}}$ & $v^{\text{ATS}}$ & $G_{\text{rel}}$\\
    \hline
    Single line & 249.32 & 219.78 & 11.85\%\\
    \hline
    Multiple lines (line 1, 3, 5) &  {152.05} & {122.51} & {19.43\%}\\
    \hline
    Multiple lines (line 19, 20, and 21)& {191.86} & {162.29} & {15.41\%}\\
    \hline
\end{tabular}
\end{table}

\subsection{Benefits of multi-hazard and multi-strategy planning}
Table \ref{Tab:multiple hazard} illustrates the benefit of incorporating multiple hazard types into the planning process. Earthquakes often lead to higher outage costs for underground lines due to extended repair times. Ignoring this risk can result in suboptimal investments. For instance, lines 13 and 15 may be selected for undergrounding based on their high load shedding costs, but their high seismic vulnerability ultimately increases total system cost. When both wind and earthquake risks are jointly considered, the model instead prioritizes alternative lines such as 20 and 21, which are less susceptible to earthquakes. This shift results in a lower overall cost. The trend is consistent across different undergrounding budget levels, demonstrating the value and cost-effectiveness of planning that accounts for multiple hazards.

\renewcommand{\arraystretch}{0.65} 
\begin{table*}[tb]
 \caption{Advantage of considering multiple hazards (MH) over single hazard (SH). The bold numbers indicate changes in the optimal line selection between SH and MH.}
 \label{Tab:multiple hazard}
 \centering
 \small
\begin{tabular}{*7c} 
 \cline{1-7}
   \multirow{2}{*}{$B^{\text{UG}}$} &\multicolumn{2}{|c}{\text{High wind and earthquake}} & \multicolumn{2}{?c}{\text{High wind only}} & \multicolumn{1}{?c} {Cost} & \multicolumn{1}{|c} {Relative}\\
 \cline{2-5}
    & \multicolumn{1}{|c}{UG line indices} & \multicolumn{1}{|c}{$v^{\text{ATS,MH}}$} &\multicolumn{1}{?c} {UG line indices} & \multicolumn{1}{|c}{$v^{\text{ATS,SH}}$ } &\multicolumn{1}{?c}{saving} &\multicolumn{1}{|c}{saving \%}\\
    \hline
   \multirow{2}{*} {35} & \multicolumn{1}{|c}{1,3,5,10,12,} & \multicolumn{1}{|c}{\multirow{2}{*}{219.78}} & \multicolumn{1}{?c}{1,3,5,10,12,} & \multicolumn{1}{|c}{ \multirow{2}{*} {473.39}}  &\multicolumn{1}{?c}{ \multirow{2}{*} {253.61}}  &\multicolumn{1}{|c}{ \multirow{2}{*}{53.57\%}}\\
   & \multicolumn{1}{|c}{18,19,\textbf{20,21}} & \multicolumn{1}{|c}{} & \multicolumn{1}{?c}{\textbf{13,15},18,19} & \multicolumn{1}{|c}{}  &\multicolumn{1}{?c}{}  &\multicolumn{1}{|c}{}\\
   \hline
   \multirow{2}{*} {43} & \multicolumn{1}{|c}{1,\textbf{2},3,5,\textbf{9},10,} & \multicolumn{1}{|c}{ \multirow{2}{*}{159.67}} & \multicolumn{1}{?c}{1,3,5,10,12,\textbf{13},} & \multicolumn{1}{|c} { \multirow{2}{*}{390.06}} &\multicolumn{1}{?c}{ \multirow{2}{*}{230.39}} &\multicolumn{1}{|c}{ \multirow{2}{*}{59.07\%}}\\
   & \multicolumn{1}{|c}{12,18,19,20,\textbf{21}} & \multicolumn{1}{|c}{} & \multicolumn{1}{?c}{\textbf{15,16},18,19,20} & \multicolumn{1}{|c} {} &\multicolumn{1}{?c}{} &\multicolumn{1}{|c}{}\\
  \hline 
\end{tabular}
\end{table*}

The advantage of integrating multiple hardening strategies, specifically undergrounding cables as a long-term measure and vegetation management as a short-term action, is demonstrated in Table \ref{Tab:multiple strategies}. Relying solely on undergrounding may leave earthquake-prone lines vulnerable to damage in their overhead segments, a risk that routine vegetation management can help mitigate. By jointly optimizing both strategies, the model captures a broader range of risks and costs, resulting in total savings of approximately \$9 million.

\renewcommand{\arraystretch}{0.6}
\begin{table}[tb]
 \caption{Advantage of using multiple strategies (MS) over single strategy (SS).}
 \label{Tab:multiple strategies}
 \centering
 \small
 \begin{tabular}{c|c|c|c|c}
 \hline
     \multirow{2}{*}{$B^{\text{UG}}$} & $v^{\text{ATS,MS}}$  & $v^{\text{ATS,SS}}$ & \text{Cost saving} & Relative  \\ 
      & $\text{(UG and VM)}$ & $\text{(only UG)}$ & $(v^{\text{ATS,SS}}-v^{\text{ATS,MS}})$ & saving \% \\
    \hline
    35 & 219.78 & 228.86 & 9.08 & 3.97\%\\
    \hline
    43 & 159.67 & 168.99 & 9.32 & 5.52\%\\
    \hline
\end{tabular}
\end{table}

\subsection{Sensitivity analysis of hardening cost, budget and line outage probability}
\begin{figure}[tb]
	\centering
	\includegraphics[scale=0.28]{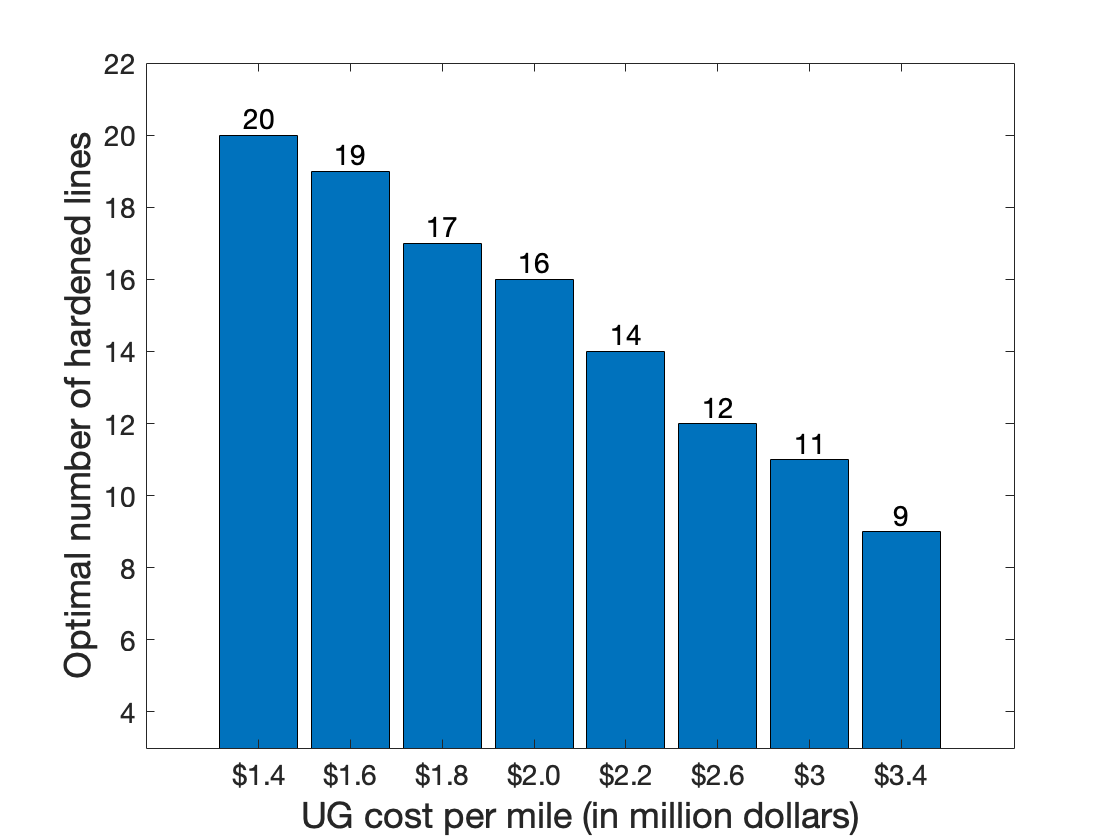}
	\caption{Sensitivity of the optimal number of hardened lines to variations in undergrounding cost $\text{IC}^\text{UG}$.}
	\label{fig:reducing ug cost}
\end{figure}

We examine how variations in the undergrounding cost per mile, denoted by $\text{IC}^{\text{UG}}$, affect the number of lines that can be hardened. Future technological advancements may reduce $\text{IC}^{\text{UG}}$, making undergrounding more economically viable. Fig. \ref{fig:reducing ug cost} illustrates this effect under a fixed undergrounding budget of \$35 million, denoted by $B^{\text{UG}}$. At the current cost of \$3.4 million per mile, the model selects nine lines for undergrounding. As $\text{IC}^{\text{UG}}$ decreases, the number of lines that can be affordably hardened increases, reaching up to 20 lines when the unit cost falls to \$1.4 million per mile.

\begin{figure}[tb]
	\centering
	\includegraphics[scale=0.28]{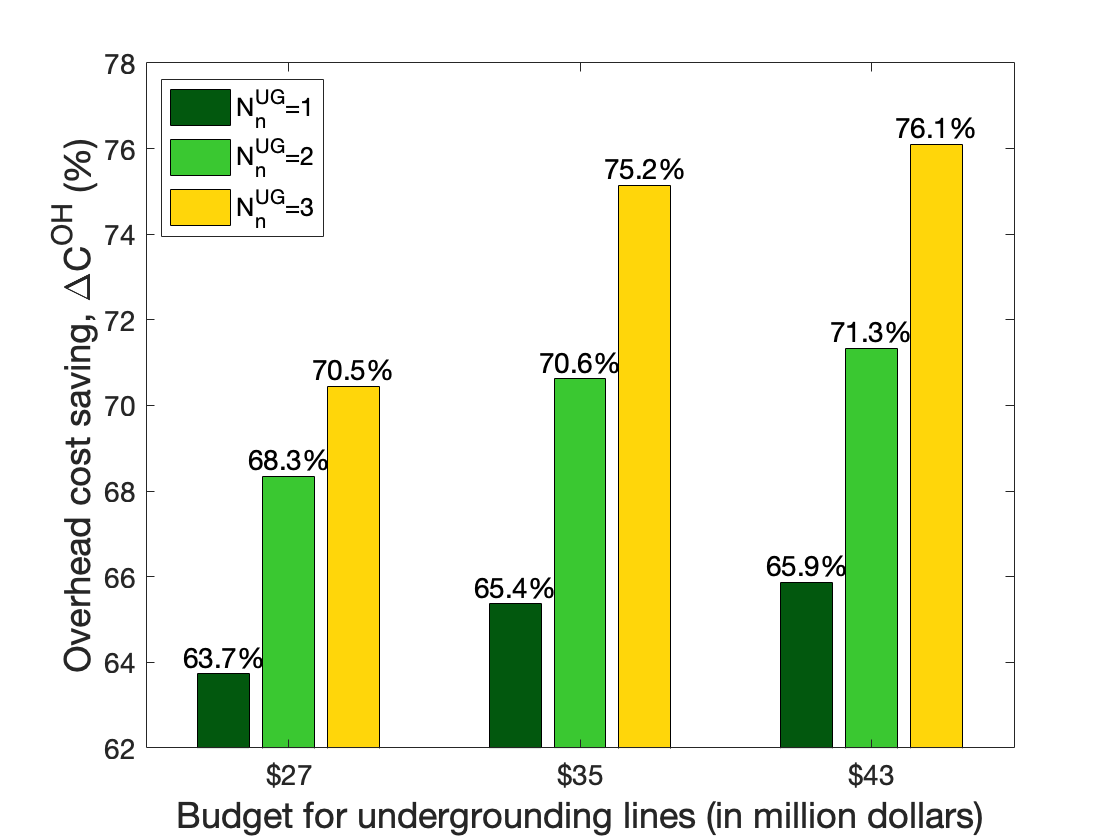}
	\caption{Sensitivity of the outage and repair cost savings for overhead lines varying undergrounding budget and $N_{n}^{\text{UG}}$.}
	\label{fig:coh saving}
\end{figure}

Fig.~\ref{fig:coh saving} illustrates the sensitivity of overhead cost savings to variations in the undergrounding budget $B^{\text{UG}}$ and the maximum number of lines that can be hardened at each node, denoted by $N_{n}^{\text{UG}}$. As expected, a higher budget allows for the undergrounding of more lines, thereby reducing the outage and repair costs $\text{C}^{\text{OH}}$,  associated with OH infrastructure. This trend holds consistently across the three different values of $N_{n}^{\text{UG}}$ considered in the analysis. Furthermore, relaxing the constraint on $N_{n}^{\text{UG}}$ leads to additional cost savings by enabling the earlier implementation of hardening strategies. The results clearly demonstrate that both a larger undergrounding budget and greater capacity for line conversion contribute significantly to minimizing outage and repair costs over the planning horizon.

Our model dynamically adapts its decisions by evaluating line outage costs associated with all hazards. We consider two wind outage probability cases ($\rho^{\text{W}}$, high and low) for line 17, which is subject to high seismic risk ($\rho_{17}^{\text{EQ}}=0.88$). Table \ref{Tab: line outage prob vs optimal decision} shows how these two risk levels govern the optimal line hardening decision. Despite having a higher load criticality factor ($L_{17}^{\text{CF}}=0.105$) compared to line 2 ($L_{2}^{\text{CF}}=0.035$), the model opts to retain line 17 overhead in a low wind risk case ($\rho_{17}^{\text{W}}=0.35$). The savings from
avoided wind-related outage and VM costs $(\text{C}^{\text{OH}})$, are insufficient to justify the expense
associated with UG. However, under an elevated wind risk ($\rho_{17}^{\text{W}}=0.75$), hardening the line leads to a notable reduction in both overhead outage and vegetation management costs compared to the previous scenario, helping to offset the expense of undergrounding. As a result, the decision to underground the line is justified, even in the presence of significant seismic risk.

\begin{table} [tb]
\centering
    \caption{Sensitivity of optimal hardening decision and cost to variations in line outage probability of line 17. The bold numbers indicate changes in the optimal line selection between two cases of wind outage probabilities.}
    \label{Tab: line outage prob vs optimal decision}
    \small
    \begin{tabular}{c|c|c|c|c}
    \hline
     $\rho_{17}^{\text{W}}$& 
     \text{Selected lines} & \multirow{2}{*}{$\alpha_{17}$} &  \multirow{2}{*} {$\text{C}^\text{OH}$} & \multirow{2}{*} {\text{VM cost}} \\ 
     ($\rho_{17}^{\text{EQ}}=0.88)$ & \text{for hardening} & &  &  \\
    \hline
    \multirow{2}{*} {0.35} & 1,\textbf{2},3,5,9,10, & \multirow{2}{*}{0} &\multirow{2}{*}{867.39} & \multirow{2}{*}{86.60}  \\
     & 12,18,19,20,21 & & & \\
    \hline
    \multirow{2}{*} {0.75} & 1,3,5,9,10,12  & \multirow{2}{*}{1} & \multirow{2}{*}{863.84} & \multirow{2}{*}{85.54} \\
     & \textbf{17},18,19,20,21 & & & \\
    \hline
\end{tabular}
\end{table}


\subsection{Performance under different scenario tree sizes and pruning methods}
Table~\ref{Tab:computation time} presents the optimal objective values and the corresponding computation times as the size of the scenario tree increases. As the number of scenarios grows, the solution becomes more cost-effective by capturing a broader range of future uncertainties, thereby supporting more informed and robust decision-making. However, this improvement is accompanied by a substantial increase in computational effort.

For example, increasing the number of scenarios from 30 to 100 reduces the cost by \$10.74 million. However, achieving a comparable reduction by increasing scenarios from 125 to 200 requires over seven times more computation time. These results indicate that beyond a certain point, the marginal benefit of additional scenarios declines, while computational cost grows substantially.

\renewcommand{\arraystretch}{0.6} 
\begin{table} [tb]
 \caption{Performance with varying number of nodes in scenario tree.}
 \label{Tab:computation time}
 \centering
 \small
 \begin{tabular}{c|c|c|c}
    \hline
    \text{Number of}& 
     \text{Total number} & \multirow{2}{*}{$v^{\text{ATS}}$} & \text{Computation  }\\ 
    \text{scenarios} & \text{of nodes} &  & \text{time} (seconds)\\
    \hline
    30 & 465 & 309.22 & 135.85\\
    \hline
    70 & 1065 & 302.55 & 475.69\\
    \hline
    100 & 1515 & 298.48 & 827.56\\
    \hline
    125 & 1890 & 295.84 & 1340.20\\
    \hline
    150 & 2265 & 292.25 & 3103.70\\
    \hline
    175 & 2640 & 288.42 & 3359.10\\
    \hline
    200 & 3015 & 285.24 & 6366.50\\
    \hline
\end{tabular}
\end{table}

In Table~\ref{Tab: BR vs FS}, we compare the optimal strategies obtained using two scenario tree pruning methods: backward reduction (BR) and forward selection (FS).  The selected lines for undergrounding are identical under both pruning methods. The relative difference between VM costs, definded as $D_{\text{rel}}=\left|\frac{\text{VM}_{\text{BR}}- \text{VM}_{\text{FS}}}{\text{VM}_{\text{BR}}}\right|\times 100\%$, is small across all budget levels. FS and BR yield similar computational performance, though FS tends to offer a modest reduction in runtime. The consistency of the results irrespective of the pruning method can be attributed in part to the assumption that line criticality factor $\text{L}^{\text{CF}}_l$ and line outage probability $\rho^{\text{W/EQ/VM}}$ are fixed all scenarios, which reduces the sensitivity of the outcomes to the choice of pruning method.

\begin{table*} [tb!]
    \caption{Comparison of optimal hardening strategies between two tree pruning methods: backward reduction (BR) and forward selection (FS). The bold numbers indicate the indices of newly added lines as the budget increases.}
    \label{Tab: BR vs FS}
    \centering
    \small
    \begin{tabular}{*7c}
    \cline{1-7}
    \multirow{3}{*}{$B^{\text{UG}}$} & \multicolumn{1}{?c}{Indices of selected lines}& \multicolumn{3}{?c}{VM cost} & \multicolumn{2}{?c}{Computation time}\\  
    & \multicolumn{1}{?c}{for undergrounding}& \multicolumn{3}{?c} {$(\$ \times 10^{3})$} & \multicolumn{2}{?c} {(seconds)}\\
    \cline{2-7}
    & \multicolumn{1}{?c} {(same for both BR and FS)} & \multicolumn{1}{?c} {BR} &  {FS} & {$D_{\text{rel}}$}& \multicolumn{1}{?c} {BR} & {FS}\\  
    \hline
    27 & \multicolumn{1}{?c}{1,3,5,10,12,18,19} & \multicolumn{1}{?c} {94.93} & 98.13 & 3.26\% & \multicolumn{1}{?c}{135.85} & 117.89\\
    \hline
    31 & \multicolumn{1}{?c}{1,3,5,10,12,18,19,\textbf{20}} & \multicolumn{1}{?c}{91.57} & 91.93 & 0.40\%& \multicolumn{1}{?c}{139.26} & 117.47\\    \hline
    35 & \multicolumn{1}{?c}{1,3,5,10,12,18,19,20,\textbf{21}} &\multicolumn{1}{?c} {89.55} & 89.75 & 0.22\% & \multicolumn{1}{?c}{144.46} & 118.29\\
    \hline
    39 & \multicolumn{1}{?c}{1,3,5,\textbf{9},10,12,18,19,20,21} & \multicolumn{1}{?c} {88.12} & 86.99 & 1.28\% & \multicolumn{1}{?c}{184.20} & 125.45\\
    \hline
    43 & \multicolumn{1}{?c}{1,\textbf{2},3,5,9,10,12,18,19,20,21} & \multicolumn{1}{?c} {86.60} & 84.93 & 1.93\% & \multicolumn{1}{?c}{141.00} & 166.82\\
    \hline
    \end{tabular}
    \end{table*}

Table~\ref{Tab:full vs pruned} presents a performance comparison between a complete scenario tree (with two branches per node) and a pruned counterpart, evaluated for time horizons $T$ from 6 to 9. The results show negligible gaps in optimal values between the two structures, even as tree size increases, demonstrating that pruning provides a computationally efficient approach to scaling stochastic optimization without compromising solution quality.

\begin{table*}[tb]
 \caption{Performance comparison between full tree and pruned tree.}
 \label{Tab:full vs pruned}
 \centering
 \footnotesize
\begin{tabular}{*{9}c} 
 \cline{1-9}
  & \multicolumn{3}{|c}{\textbf{Full tree}} & \multicolumn{3}{?c}{\textbf{Pruned tree}} & \multicolumn{1}{?c} {Relative} & \multicolumn{1}{|c} {Relative}\\
 \cline{1-7}
  \multirow{2}{*}{$T$} &\multicolumn{1}{|c}{No. of} &\multicolumn{1}{c}{\multirow{2}{*}{$v^{\text{ATS}}$}} &\multicolumn{1}{c}{Comp.} & \multicolumn{1}{?c}{No. of} & \multicolumn{1}{c}{\multirow{2}{*}{$v^{\text{ATS}}$}} &\multicolumn{1}{c}{Comp.} &\multicolumn{1}{?c}{optimality} & \multicolumn{1}{|c} {time}\\
    & \multicolumn{1}{|c}{nodes} &  \multicolumn{1}{c}{}&\multicolumn{1}{c}{time (sec.)} & \multicolumn{1}{?c}{nodes}&  \multicolumn{1}{c}{}&\multicolumn{1}{c}{time (sec.)} &\multicolumn{1}{?c}{gap} & \multicolumn{1}{|c} {reduction}\\
    \hline
  6 &\multicolumn{1}{|c}{63} & \multicolumn{1}{c}{109.59} & \multicolumn{1} {c} {5.53} & \multicolumn{1}{?c}{63} & \multicolumn{1}{c}{104.93} & \multicolumn{1} {c}{4.29} &\multicolumn{1}{?c}{4.25\%} &\multicolumn{1}{|c}{22.42\%}\\
  7 &\multicolumn{1}{|c}{127} & \multicolumn{1}{c}{107.54} & \multicolumn{1}{c}{11.17} &\multicolumn{1}{?c}{126} & \multicolumn{1}{c}{113.45} & \multicolumn{1}{c}{9.26} &\multicolumn{1}{?c}{5.50\%} &\multicolumn{1}{|c}{17.10\%}\\
  8 &\multicolumn{1}{|c}{255} & \multicolumn{1}{c}{102.03} & \multicolumn{1}{c}{40.87} & \multicolumn{1}{?c}{256} & \multicolumn{1}{c}{98.25} & \multicolumn{1}{c}{25.81} &\multicolumn{1}{?c}{3.71\%} &\multicolumn{1}{|c}{36.85\%}\\
  9 &\multicolumn{1}{|c}{511} & \multicolumn{1}{c}{93.54} & \multicolumn{1}{c}{250.51} & \multicolumn{1}{?c}{513} & \multicolumn{1}{c}{92.84} & \multicolumn{1}{c}{87.86} &\multicolumn{1}{?c}{0.75\%} &\multicolumn{1}{|c}{64.93\%}\\
  \hline 
\end{tabular}
\end{table*}

To evaluate scalability, in Table \ref{Tab: comp time comparison2} we extend our analysis to the IEEE 69-bus system with heterogeneous line lengths, and compare its performance against the IEEE 22-bus system with uniform line lengths. As expected, computation time increases with both system size and line heterogeneity. Nevertheless, the model remains computationally tractable across a range of budget levels and scenario tree sizes. Finally, while computational efficiency is desirable, it is not a limiting factor in the context of long term planning, which is typically conducted offline and is not subject to real time constraints. Combined with the ongoing advances in high-performance computing, these results support the scalability and practical applicability of our framework in a real-world power grid.

\begin{table*} [tb]
\centering
        \caption{Computation time under varying grid sizes and line configurations: uniform lengths (UL) vs. heterogeneous lengths (HL).}
        \label{Tab: comp time comparison2}
        \centering
        \small
            \begin{tabular}{*5c}
                \hline
                 \multirow{2}{*}{$B^{\text{UG}}$}&\multicolumn{1}{?c}{Number of}&\multicolumn{1}{?c}{Number of}
                  & \multicolumn{2}{?c}{Computation time (seconds)} \\ 
                  \cline{4-5}
                 &\multicolumn{1}{?c}{scenarios} &\multicolumn{1}{?c}{nodes} & \multicolumn{1}{?c}{22-bus with UL}& 69-bus with HL\\
                 \hline
                 \multirow{3}{*} {27} & \multicolumn{1}{?c}{30} &\multicolumn{1}{?c}{465} & \multicolumn{1}{?c}{135.85} & 802.05\\
                 & \multicolumn{1}{?c}{70} & \multicolumn{1}{?c}{1065} & \multicolumn{1}{?c}{475.69}& 3240.70\\
                 & \multicolumn{1}{?c}{100} & \multicolumn{1}{?c}{1515} & \multicolumn{1}{?c}{827.56}& 8217.70\\
                 \hline
                 \multirow{3}{*} {43} & \multicolumn{1}{?c}{30} & \multicolumn{1}{?c} {465} & \multicolumn{1}{?c}{141.00} & 701.24\\
                & \multicolumn{1}{?c}{70} & \multicolumn{1}{?c}{1065} & \multicolumn{1}{?c}{588.00 }& 3725.50\\
                & \multicolumn{1}{?c}{100} & \multicolumn{1}{?c}{1515} & \multicolumn{1}{?c}{1270.80}& 7233.70\\
                \hline
                 \multirow{3}{*} {63} & \multicolumn{1}{?c}{30} & \multicolumn{1}{?c} {465} & \multicolumn{1}{?c}{171.00} & 688.87\\
                 & \multicolumn{1}{?c}{70} & \multicolumn{1}{?c}{1065} & \multicolumn{1}{?c}{947.73}& 4767.00\\
                 & \multicolumn{1}{?c}{100} & \multicolumn{1}{?c}{1515} & \multicolumn{1}{?c}{2111.00}& 7804.10\\
                 \hline
        \end{tabular}
    \end{table*}

\section{Conclusion}
\label{conclusion}

In this work, we propose an optimization framework that integrates two complementary hardening strategies, undergrounding (UG) and vegetation management (VM), to enhance the resilience of distribution networks. The model is solved using an adaptive two-stage stochastic programming approach, which allows for a single revision of early-stage decisions based on the unfolding of uncertain future events. Our results show that jointly optimizing UG and VM while accounting for multiple hazards and associated investment, outage, and repair costs leads to significant cost savings and improved grid resilience. In addition, our study demonstrates the advantages of adopting the adaptive two-stage framework over conventional two-stage models. These benefits are particularly evident under broader investment budgets and increased uncertainty, where the ability to revise decisions in response to emerging scenarios provides considerable value. 

Planning a resilient grid hardening strategy is a complex task influenced by utilities' risk preferences, geographical location, hazard exposure, and various cost
parameters. Risk-averse utilities may prioritize extensive hardening to reduce outage exposure, while risk-tolerant ones may adopt more
selective hardening for highly critical lines, relying more on operational measures
such as vegetation management. Regional hazard profiles and economic trade-offs between various cost components
further shape these decisions. Gaining deeper insight into the interplay of risk attitude, asset vulnerabilities, and cost dynamics can guide utility investments and offers a promising direction for future research we aim to pursue.


\bibliographystyle{IEEEtran}
\bibliography{refs,IEEEabrv}

@BOOK{Molzahn-Hiskens-2019,
  author={Molzahn, Daniel and Hiskens, Ian},
  title={A Survey of Relaxations and Approximations of the Power Flow Equations},
  publisher={Now Publishers},
  year={2019}}

@ARTICLE{Masoud-Low-2013,
  author={Farivar, Masoud and Low, Steven H.},
  journal={IEEE Transactions on Power Systems}, 
  title={Branch Flow Model: Relaxations and Convexification—Part {I}}, 
  year={2013},
  volume={28},
  number={3},
  pages={2554-2564}
}

@ARTICLE{number_projection,
    author = {Smith, Adam B.},
    journal={},
    title = {{U.S.} Billion-Dollar Weather and Climate Disasters, 1980 - present.},
    institution = {{NOAA} National Centers for Environmental Information {(NCEI)}},
    month={Jun.},
    year = {2024},
    volume={},
    number={},
    pages={},
    howpublished = {https://www.ncei.noaa.gov/access/billions/},
    doi={10.25921/stkw-7w73}
}

@INPROCEEDINGS{distribution_outage_duration,
  author={Zapata, C. J. and Silva, S. C. and Burbano, O. L.},
  booktitle={2008 IEEE/PES T\&D Conf. and Expo.: Latin America}, 
  title={Repair models of power distribution components}, 
  year={2008},
  volume={},
  number={},
  pages={1-6},
  doi={10.1109/TDC-LA.2008.4641851}}

@INPROCEEDINGS{hurricane1,
  author={Poudyal, Abodh and Dubey, Anamika and Iyengar, Vishnu and Garcia-Camargo, Diego},
  booktitle={2022 IEEE Power \& Energy Soc. Gen. Meeting (PESGM)}, 
  title={Spatiotemporal Impact Assessment of Hurricanes on Electric Power Systems}, 
  year={2022},
  volume={},
  number={},
  pages={1-5},
  doi={10.1109/PESGM48719.2022.9917119}}

@inproceedings{flood2,
  title={Storm \& flood hardening of electrical substations},
  author={Boggess, JM and Becker, GW and Mitchell, MK},
  booktitle={2014 IEEE/PES T\&D Conf. and Expo.},
  pages={1--5},
  year={2014},
}

@INPROCEEDINGS{tri-levelopt,
  author={Ahmadi, Mehdi and Bahrami, Mahdi and Vakilian, Mehdi and Lehtonen, Matti},
  booktitle={2020 IEEE PES T\&D Conf. and Expo.- Latin America}, 
  title={Application of Hardening Strategies and {DG} Placement to Improve Distribution Network Resilience against Earthquakes}, 
  year={2020},
  volume={},
  number={},
  pages={1-6},
  doi={10.1109/TDLA47668.2020.9326205}}

@ARTICLE{ess_allocation,
  author={Nazemi, Mostafa and Moeini-Aghtaie, Moein and Fotuhi-Firuzabad, Mahmud and Dehghanian, Payman},
  journal={{IEEE} Trans. Sustainable Energy}, 
  title={Energy Storage Planning for Enhanced Resilience of Power Distribution Networks Against Earthquakes}, 
  year={2020},
  volume={11},
  number={2},
  pages={795-806},
  doi={10.1109/TSTE.2019.2907613}}

@misc{multi-strategies,
      title={Resilience-driven Planning of Electric Power Systems Against Extreme Weather Events}, 
      author={Abodh Poudyal and Shishir Lamichhane and Anamika Dubey and Josue Campos do Prado},
      year={2023},
      eprint={2311.15117},
      archivePrefix={arXiv},
      primaryClass={eess.SY},
      url={https://arxiv.org/abs/2311.15117}, 
}

@ARTICLE{UG1,
  author={Trakas, Dimitris N. and Hatziargyriou, Nikos D.},
  journal={{IEEE} Trans. Power Syst}, 
  title={Strengthening Transmission System Resilience Against Extreme Weather Events by Undergrounding Selected Lines}, 
  year={2022},
  volume={37},
  number={4},
  pages={2808-2820},
  doi={10.1109/TPWRS.2021.3128020}}

@ARTICLE{DRO2,
  author={Zhao, Pengfei and Gu, Chenghong and Cao, Zhidong and Shen, Yichen and Teng, Fei and Chen, Xinlei and Wu, Chenye and Huo, Da and Xu, Xu and Li, Shuangqi},
  journal={{IEEE} Trans Ind. Informat}, 
  title={Data-Driven Multi-Energy Investment and Management Under Earthquakes}, 
  year={2021},
  volume={17},
  number={10},
  pages={6939-6950},
  doi={10.1109/TII.2020.3043086}}

@ARTICLE{ATSBestie,
  author={Beste Basciftci and Shabbir Ahmed and Nagi Gebraeel},
  journal={Manuf. Serv. Oper. Manag.}, 
  title={Adaptive Two-stage Stochastic Programming with an Analysis on Capacity Expansion Planning Problem}, 
  year={2024},
  volume={26},
  number={6},
  pages={2121-2141},
  doi={10.1287/msom.2023.0157}}

@ARTICLE{ATSShi,
  author={Shi, Wenlong and Liang, Hao and Bittner, Myrna},
  journal={{IEEE} Trans Ind. Informat}, 
  title={Data-Driven Resilience Enhancement for Power Distribution Systems Against Multishocks of Earthquakes}, 
  year={2024},
  volume={20},
  number={5},
  pages={7357-7369},
  doi={10.1109/TII.2024.3359437}}

@article{dupavcova2000scenarios,
  title={Scenarios for multistage stochastic programs},
  author={Dupa{\v{c}}ov{\'a}, Jitka and Consigli, Giorgio and Wallace, Stein W},
  journal={Ann. Oper. Res.},
  volume={100},
  number={1},
  pages={25--53},
  year={2000},
  publisher={Springer}
}

@inproceedings{scenred,
  title={Scenario reduction and scenario tree construction for power management problems},
  author={Growe-Kuska, Nicole and Heitsch, Holger and Romisch, Werner},
  booktitle={2003 IEEE Bologna Power Tech Conference Proceedings},
  volume={3},
  year={2003},
  organization={IEEE}
}

@inproceedings{multistage,
  title={A multistage stochastic programming model for hurricane relief logistics planning},
  author={Siddig, Murwan and Song, Yongjia},
  booktitle={Proc. 2021 IISE Annu. Conf.},
  year={2021}
}

@book{shapiro2003,
  title={Stochastic programming},
  author={Shapiro, Alexander and Andrzej, P Ruszczy{\'L}},
  volume={12},
  year={2003},
  publisher={Elsevier Amsterdam, Netherlands}
}

@article{FSandBR,
  title={Scenario reduction algorithms in stochastic programming},
  author={Heitsch, Holger and R{\"o}misch, Werner},
  journal={Comput. Optim. Appl.},
  volume={24},
  pages={187--206},
  year={2003},
  publisher={Springer}
}

@article{robust,
  title={Robust expansion planning and hardening strategy of meshed multi-energy distribution networks for resilience enhancement},
  author={Li, Tingjun and Han, Xiaoqing and Wu, Wenchuan and Sun, Hongbin},
  journal={Appl. Energy},
  volume={341},
  pages={121066},
  year={2023},
  publisher={Elsevier}
}

@article{AE_multiple,
  title={Resilience enhancement of distribution network under typhoon disaster based on two-stage stochastic programming},
  author={Hou, Hui and Tang, Junyi and Zhang, Zhiwei and Wang, Zhuo and Wei, Ruizeng and Wang, Lei and He, Huan and Wu, Xixiu},
  journal={Appl. Energy},
  volume={338},
  pages={120892},
  year={2023},
  publisher={Elsevier}
}

@article{stochastic_DR,
  title={Resilience-oriented planning of integrated electricity and heat systems: A stochastic distributionally robust optimization approach},
  author={Zhou, Yizhou and Li, Xiang and Han, Haiteng and Wei, Zhinong and Zang, Haixiang and Sun, Guoqiang and Chen, Sheng},
  journal={Appl. Energy},
  volume={353},
  pages={122053},
  year={2024},
  publisher={Elsevier}
}

@techreport{financial_loss,
    title = {Extreme Weather Costs the {U.S.} Billions Each Year — Is Your Facility at Risk?},
    institution = {Unison Energy},
    month={May},
    year = {2023},
    URL={https://unisonenergy.com/resources/blog/extreme-weather-cost/}
}

@article{ess_dg_allocation,
  title={Strengthening distribution systems after earthquakes with a new analytical model},
  author={Akdag, Ozan},
  journal={Electr. Power. Syst. Res.},
  volume={232},
  pages={110337},
  year={2024},
  publisher={Elsevier}
}

@article{vm,
  title={Assessing the effects of a vegetation management standard on distribution grid outage rates},
  author={Cerrai, Diego and Watson, Peter and Anagnostou, Emmanouil N},
  journal={Electr. Power. Syst. Res.},
  volume={175},
  pages={105909},
  year={2019},
  publisher={Elsevier}
}

@article{cc_influence_review,
  title={Influence of extreme weather and climate change on the resilience of power systems: Impacts and possible mitigation strategies},
  author={Panteli, Mathaios and Mancarella, Pierluigi},
  journal={Electr. Power. Syst. Res.},
  volume={127},
  pages={259--270},
  year={2015},
  publisher={Elsevier}
}

@article{wildfire3,
  title={A framework for risk assessment and optimal line upgrade selection to mitigate wildfire risk},
  author={Taylor, Sofia and Roald, Line A},
  journal={Electr. Power. Syst. Res.},
  volume={213},
  pages={108592},
  year={2022},
  publisher={Elsevier}
}

@article{veg_risk,
  title={Modeling the impact of local environmental variables on tree-related power outages along distribution powerlines},
  author={Wedagedara, Harshana and Witharana, Chandi and Fahey, Robert and Cerrai, Diego and Joshi, Durga and Parent, Jason},
  journal={Electr. Power. Syst. Res.},
  volume={221},
  pages={109486},
  year={2023},
  publisher={Elsevier}
}

@article{multi2,
  title={Improving distribution system resilience by undergrounding lines and deploying mobile generators},
  author={Taheri, Babak and Molzahn, Daniel K and Grijalva, Santiago},
  journal={Electr. Power. Syst. Res.},
  volume={214},
  pages={108804},
  year={2023},
  publisher={Elsevier}
}

@article{hurricane2,
  title={Aggregated vulnerability assessment of power transmission lines under operational and hurricane induced outages},
  author={Khan, Saad Ullah and Khan, Muhammad Sajid and Farooq, Hamza},
  journal={Electr. Power. Syst. Res.},
  volume={240},
  pages={111262},
  year={2025},
  publisher={Elsevier}
}

@article{eq3,
  title={Rapid assessment of substation earthquake risk based on minimal cut sets},
  author={Liu, Xiao and Xie, Qiang and Zhu, Wang},
  journal={Electr. Power. Syst. Res.},
  volume={229},
  pages={110175},
  year={2024},
  publisher={Elsevier}
}

\end{document}